\documentclass[11pt]{article}
\pdfoutput=1
\usepackage{graphicx}
\usepackage{epsfig}
\usepackage{amsmath}
\usepackage{accents}
\usepackage{amsthm}
\usepackage{amssymb}
\usepackage{multirow}
\usepackage{color}
\usepackage[nosort]{cite}
\usepackage{hyperref}
\usepackage[fontsize=11.35pt]{scrextend}
\usepackage{braket,mathrsfs,slashed}
\usepackage{float}
\usepackage{setspace}
\usepackage[caption=false]{subfig}
\newlength{\abstractwidth}
\newcommand{\be}{\begin{equation}}
\newcommand{\ee}{\end{equation}}

\renewcommand{\title}[1]{\vbox{\center\bf{\Large{#1}}}\vspace{5mm}}
\renewcommand{\author}[1]{\vbox{\center#1}\vspace{5mm}}
\newcommand{\address}[1]{\vbox{\center\em#1}}
\newcommand{\email}[1]{\vbox{\center\tt#1}\vspace{5mm}}

\usepackage{dsfont}

\usepackage{rotating}

\usepackage{dsfont}
\usepackage[paper=a4paper,margin=1.5cm]{geometry}
\usepackage[utf8]{inputenc}
\usepackage{rotating}
\usepackage{soul}

\usepackage{mathrsfs} 
\usepackage{bbm}
\usepackage{etoolbox} 
\usepackage{multirow}

\renewcommand\[{\begin{equation}}
\renewcommand\]{\end{equation}}

\newcommand{\ba}{\begin{eqnarray}}
\newcommand{\ea}{\end{eqnarray}}

\hypersetup{pdfstartview=FitV,colorlinks=true,linkcolor=midblue,citecolor=midblue,filecolor=midblue,urlcolor=midblue}

\definecolor{midblue}{rgb}{0,0,0.6}

\begin{document}
	
\newgeometry{top=3.1cm,bottom=3.1cm,right=2.4cm,left=2.4cm}

	\begin{titlepage}
	\begin{center}
		\hfill \\
		\vskip 0.5cm

		\title{Asymptotic Quantum Dynamics of Ghost Fields}

		\author{\large Luca Buoninfante}
		
		\address{Departamento de Física de Partículas,\\ Instituto Galego de Física de Altas Enerxías (IGFAE),\\  Universidade de Santiago de Compostela, Spain}
		\email{\rm \href{mailto:luca.buoninfante@usc.es}{luca.buoninfante@usc.es}}

	\end{center}

\begin{abstract}
The dressed propagator of a ghost coupled to ordinary fields develops a pair of complex conjugate poles in the first Riemann sheet above the multi-particle threshold. We study the implications of this pole structure for the asymptotic field and its negative-norm one-particle state. Within the operator formalism of local quantum field theory, we show that certain quadratic couplings between the ghost field and the composite field of the multi-particle state persist at asymptotic times. These induce quantum interference effects that render the negative-norm one-particle state non-orthogonal to a superposition of positive-norm multi-particle states. Consequently, no free asymptotic one-particle ghost state exists, while the true free asymptotic states have zero norm and hence vanishing localization probability. The real and imaginary parts of the complex mass admit a clear physical interpretation; in particular, the inverse imaginary part sets the timescale for the onset of non-orthogonality. A~freely propagating ghost is therefore confined to time intervals much shorter than its inverse width, so that a detector can never observe an isolated ghost particle asymptotically. Open questions and potential applications are discussed in the conclusions.
\end{abstract}

\end{titlepage}

{
	\hypersetup{linkcolor=black}
	\tableofcontents
}

\baselineskip=17.63pt



\newpage

\section{Introduction}

Ghosts are fields whose kinetic term has opposite sign to that of ordinary fields. This sign difference has important mathematical and physical consequences. First, it enters the numerator of the propagator, enabling cancellations that can improve theory's ultraviolet behavior, as in four-derivative theories where the propagator scales as $1/p^4$ at high energies due to a massive ghost component~\cite{Pais:1950za,Bender:2007wu,Salvio:2015gsi,Holdom:2023usn,Holdom:2024cfq}. Key examples include Lee-Wick models~\cite{Lee:1969fy,Lee:1970iw,Lee:chicago,Cutkosky:1969fq} and quadratic gravity~\cite{Stelle:1976gc,Tomboulis:1980bs,Avramidi:1985ki,Salvio:2018crh,Anselmi:2018tmf,Donoghue:2021cza,Holdom:2021hlo,Buoninfante:2023ryt,Buoninfante:2025dgy,Kuntz:2024rzu,Oda:2025buc,Kumar:2026qnw}. Second, the minus sign enters the denominator of the dressed propagator, giving rise to complex conjugate poles in the first Riemann sheet above the multi-particle threshold~\cite{Coleman:1969xz,Kubo:2024ysu,Buoninfante:2025klm}.

The peculiar pole structure of the dressed ghost propagator has raised concerns and prompted different approaches over time. In Lee-Wick models, complex poles were initially interpreted as signaling that ghosts decay and disappear from the asymptotic spectrum~\cite{Lee:1969fy,Lee:1970iw,Lee:chicago}, an idea later revisited in~\cite{Grinstein:2008bg,Donoghue:2019fcb}. An alternative approach is to introduce new diagrammatic rules so that such ghost states never go on-shell but remain purely virtual~\cite{Anselmi:2021hab}. On the other hand, by working within the operator formalism of local quantum field theory~(QFT), one can show that first-sheet complex conjugate poles imply that ghosts cannot decay and, in particular, cannot be removed from the set of asymptotic states~\cite{Kubo:2024ysu}. This fact has been viewed as problematic, as it may imply that negative probabilities associated with ghosts are observable at asymptotic times~\cite{Kubo:2023lpz}.

This concern would be justified if a freely propagating ghost existed asymptotically. Our aim, however, is to demonstrate the opposite within the operator formalism of~local~QFT. In~particular, we will show that nontrivial interactions between the ghost field and the composite field of the multi-particle state persist at asymptotic times, inducing quantum interference effects that render the negative-norm one-particle state non-orthogonal to a superposition of positive-norm multi-particle states. As a result, no free asymptotic one-particle ghost state exists.

The paper is organized as follows.
\vspace{-2mm}
\begin{description}
	
	\item[Sec.~\ref{sec:field-theory-model}:] We introduce a field-theory model and review general properties of the dressed ghost propagator, including its pole structure, spectral representation, and a sum rule.
	\vspace{-2mm}
	\item[Sec.~\ref{sec:asympt-fields}:] We show that a local and Hermitian Lagrangian description reproducing the complex conjugate pole structure requires effectively doubling the asymptotic ghost field by introducing an additional field, identified as the asymptotic limit of the composite field of the multi-particle state. Their interactions render the ghost never free asymptotically.
	\vspace{-2mm}
	\item[Secs.~\ref{sec:compl-quant-eff},~\ref{sec:phys-impl}:] We discuss complementary quantum dynamical effects and provide a physical interpretation of the real and (inverse) imaginary parts of the complex mass in terms of physical real mass, interaction coupling, and timescale.
	\vspace{-2mm}
	\item[Sec.~\ref{sec:concl-outl}:] We summarize the main results, highlight several open questions, and discuss potential future applications.

\end{description}

Throughout the paper we work in Natural units ($\hbar=1=c$), in $3+1$ spacetime dimensions, and adopt the mostly plus convention for the metric signature $(-,+,+,+)$.

\section{Field-theory model}\label{sec:field-theory-model}

Let us consider the field-theory model described by the Lagrangian density
\begin{equation}
	\mathcal{L}_{\chi\phi}= -\frac{1}{2}\left(\partial_\mu\chi\partial^\mu\chi +\mu^2\chi^2\right)+\frac{1}{2} \left(\partial_\mu\phi\partial^\mu \phi+m^2\phi^2\right)-\frac{g}{2} \phi\chi^2\,,
	\label{lagrangian}
\end{equation}
where $\chi(x)$ is an ordinary scalar field of mass $\mu>0$, while $\phi(x)$ is a ghost field due to the opposite sign of its kinetic term, of mass $m>0$. The two masses are assumed to be already renormalized, consistently with the renormalization condition imposed below.

The free propagators of the two fields are
\begin{equation}
	G_\chi(-p^2) = \frac{-i}{p^2+\mu^2-i\epsilon} \,,	\qquad 
	G_{\phi}(-p^2)= \frac{+i}{p^2+m^2-i\epsilon}\,,
	\label{propagator-phi-free}
\end{equation}
where $\epsilon\rightarrow 0^+$ is understood. The choice of Feynman prescription for the $\phi$ propagator implies that the ghost field is quantized in the perturbative QFT framework in a way compatible with a bounded Hamiltonian, the unitarity of the evolution operator, and an indefinite-norm vector space: states with an odd (even) number of ghost particles have negative (positive) norm~\cite{Buoninfante:2025klm}.

Despite the simplicity of the Lagrangian~\eqref{lagrangian}, the results obtained in this work are quite general and extend, for example, to four-derivative theories~\cite{Pais:1950za,Bender:2007wu,Salvio:2015gsi,Holdom:2023usn,Holdom:2024cfq}, with relevant implications for Lee-Wick models~\cite{Lee:1969fy,Lee:1970iw,Lee:chicago,Cutkosky:1969fq} and quadratic gravity~\cite{Stelle:1976gc,Tomboulis:1980bs,Avramidi:1985ki,Salvio:2018crh,Anselmi:2018tmf,Donoghue:2021cza,Holdom:2021hlo,Buoninfante:2023ryt,Buoninfante:2025dgy,Kuntz:2024rzu,Oda:2025buc,Kumar:2026qnw}, as discussed in the conclusions.

\subsection{Dressed ghost propagator}

The interaction term $g\phi\chi^2/2$ induces quantum corrections in the $\phi$ propagator that can be computed by resumming the self-energies to all orders in perturbation theory. For the purposes of our discussion, it is sufficient to focus on the one-loop self-energy, corresponding to the bubble diagram with $\chi$ propagators on the internal lines, which as a function of $-p^2\rightarrow z\in\mathbb{C}$ reads
\begin{eqnarray}
		i\Sigma(z)\!\!\!&=&\!\!\! \frac{(-ig)^2}{2}\int \frac{{\rm d}^4k}{(2\pi)^4} G_\chi\left(-(k-p)^2\right)G_\chi\left(-k^2\right)\nonumber\\
		\!\!\!&=&\!\!\!\,i\frac{g^2}{32\pi^2}\left[2-\log\left(\frac{\mu^2}{\Lambda^2}\right)-2\sqrt{\frac{4\mu^2-z}{z}}\arctan\sqrt{\frac{z}{4\mu^2-z}}\right]\,,
	\label{self-energy}
\end{eqnarray}
where $\Lambda$ is the renormalization scale in cut-off regularization. For a real and time-like momentum squared, i.e. $z \to -p^2 > 0$, the real and imaginary parts of the self-energy are
\begin{equation}
	{\rm Re}\left[\Sigma(-p^2)\right]=\frac{g^2}{32\pi^2}\left[ 2-\log\left(\frac{\mu^2}{\Lambda^2}\right)+\sqrt{1+\frac{4\mu^2}{p^2}}\log\left(\frac{1-\sqrt{1+4\mu^2/p^2}}{1+\sqrt{1+4\mu^2/p^2}}\right) \right]
	\label{real-self-energy}
\end{equation}
and
\begin{equation}
	{\rm Im}\left[\Sigma(-p^2\pm i\epsilon)\right]=\pm\frac{g^2}{32\pi}\sqrt{1+\frac{4\mu^2}{p^2}}\,\theta\left(-p^2-4\mu^2\right)\,.
	\label{imag-self-en}
\end{equation}
It is worth mentioning that the expression~\eqref{imag-self-en} refers to the imaginary part of the self-energy on the first Riemann sheet of the complex $z$ plane. The signs would be opposite if the imaginary part were evaluated on the second sheet~\cite{Buoninfante:2025klm}.

The dressed $\phi$ propagator $\bar{G}_{\phi}(z)$, as a function of the complex momentum $-p^2 \to z \in \mathbb{C}$, can be computed by resumming the following geometric series:
\begin{equation}
	\bar{G}_{\phi}(z) =  G_{\phi}(z)\sum_{n=0}^\infty \left[i\Sigma(z)G_{\phi}(z) \right]^n=\frac{G_\phi(z)}{1-i\Sigma(z)G_\phi(z)}=\frac{-i}{z-m^2-\Sigma(z)}\,.
	\label{dressed-propagator}
\end{equation}
If $\phi$ were an ordinary field, the dressed propagator would have no poles in the first Riemann sheet of the complex $z$ plane above the multi-particle threshold $(m>2\mu)$, but would instead exhibit a pair of complex conjugate poles in the second sheet. Physically, this means that a positive-norm one-particle state can decay and disappear from the set of asymptotic states. In contrast, if $\phi$ is a ghost, as in our case, the dressed propagator acquires a pair of complex conjugate poles in the first sheet above the multi-particle threshold, namely there exists a $z_{\rm pole}=M^2\in\mathbb{C}^{(\text{I sheet})}$~such~that
\begin{equation}
i\,\Sigma (M^2)G_\phi(M^2)=1\,,\qquad i\,\Sigma(M^{*2})G_\phi(M^{*2})=1\,,
\label{poles-zp}
\end{equation}
where we have used the reflection properties $\Sigma(z^*)=\Sigma^*(z)$ and $G_\phi(z^*)=-G_\phi^*(z)$. 

This striking difference between the ordinary and ghost cases is due to the fact that the opposite sign in the numerator of the free ghost propagator in~\eqref{propagator-phi-free} also affects the sign in front of the self-energy in the denominator of~\eqref{dressed-propagator}. Therefore, above the multi-particle threshold, the analyticity domain of the dressed ghost propagator in the first sheet of the complex $z$ plane is $\mathbb{C}^{(\text{I sheet})}\setminus \lbrace M^2, M^{*2}\rbrace$, up to the branch cut along $[4\mu^2,+\infty)$~\cite{Kubo:2024ysu,Buoninfante:2025klm}.

The complex conjugate poles can be written as $M^2=m^2+im\Gamma$ and $M^{*2}=m^2-im\Gamma$, upon imposing the renormalization condition ${\rm Re}[\Sigma(m^2\pm im\Gamma)]=0$, where the imaginary part solves the poles equation
\begin{eqnarray}
\pm i m\Gamma =i {\rm Im}[\Sigma(m^2\pm im\Gamma)]\,.
\end{eqnarray}
In the narrow-width approximation $\Gamma/m\ll 1$, using~\eqref{imag-self-en} with $-p^2\pm i\epsilon$ replaced by $m^2\pm im\Gamma$, we obtain an explicit expression for the imaginary part of the poles at order $\mathcal{O}(g^2)$:
\begin{eqnarray}
\pm im\Gamma \simeq \pm i \frac{g^2}{32\pi}\sqrt{1-\frac{4\mu^2}{m^2}}\qquad \Leftrightarrow \qquad \Gamma\simeq  \frac{g^2}{32\pi m} \sqrt{1-\frac{4\mu^2}{m^2}}\,.
\end{eqnarray}

\begin{figure}[t!]
\centering
\includegraphics[scale=0.42]{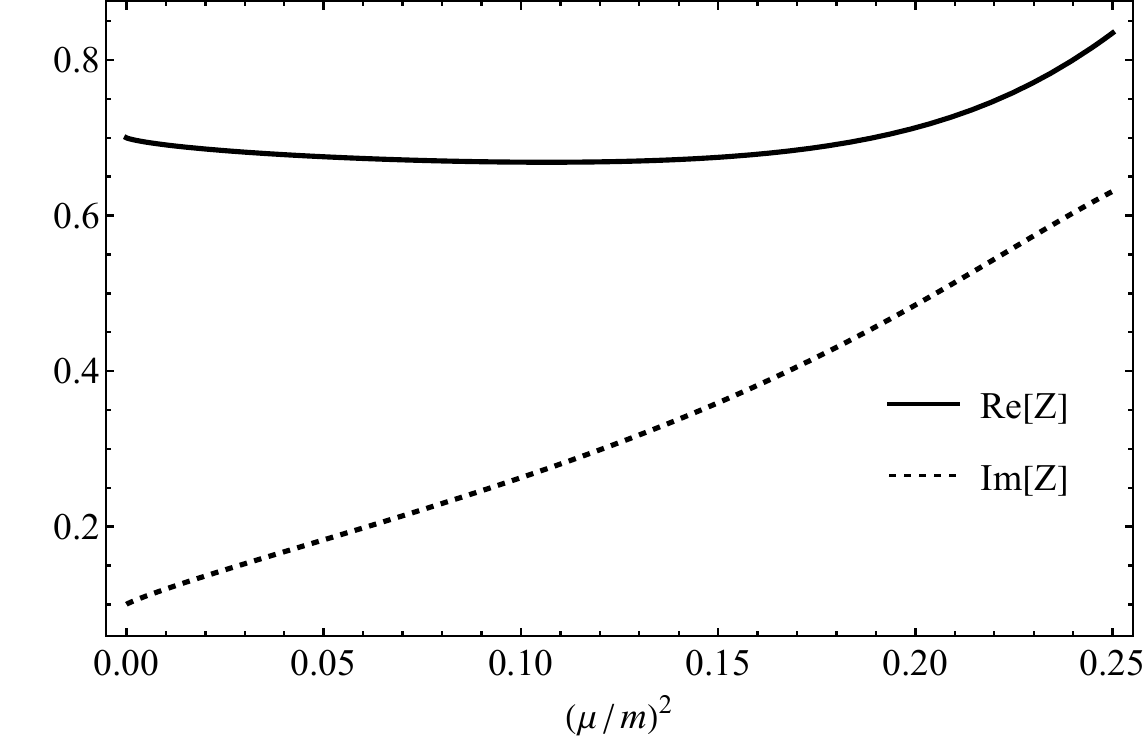}
\protect\caption{Real (solid line) and imaginary (dashed line) parts of the wave-function renormalization constant $Z={\rm Re}[Z]+i{\rm Im}[Z]$ are plotted as functions of $(\mu/m)^2$. For illustrative purposes we set $\Gamma/m=0.5$ and  $g^2/(32\pi^2m^2)=0.5$. Since we are working above the multi-particle threshold, the region of interest is $(\mu/m)^2<0.25$. We find ${\rm Re}[Z]>0$ and ${\rm Im}[Z]>0.$ }
\label{fig1}
\end{figure}

The residue of the dressed propagator at $M^2$ is given by $-iZ$, where $Z \in \mathbb{C}$ is a complex wave-function renormalization constant:
\begin{eqnarray}
Z^{-1}=1- \left.\text{Re}\left[\frac{{\rm d}\Sigma(z)}{{\rm d}z}\right]\right|_{z=M^2} -  i \left.\text{Im}\left[ \frac{{\rm d}\Sigma(z)}{{\rm d}z}\right]\right|_{z=M^2}\,.
\end{eqnarray}
Consequently, the residue at $M^{*2}$ is $-iZ^*.$ Both the real and imaginary parts of $Z$ can be shown to be positive in perturbation theory ($g^2/(32\pi^2 m^2)<1$). In Fig.~\ref{fig1} we plot them as functions of $(\mu/m)^2$, keeping $\Gamma/m$ and $g^2/(32\pi^2 m^2)$ fixed. Approximate expressions can instead be obtained for $\mu/m\ll 1$, $\Gamma/m\ll 1$: ${\rm Re}[Z]\simeq 1 - g^2/(32\pi^2 m^2) > 0$ and ${\rm Im}[Z]\simeq g^2\Gamma/(32\pi^2 m^3) > 0$. Note that in this regime we have $g^2/(32\pi^2 m^2)\simeq \Gamma/m$, so that ${\rm Im}[Z]\simeq (\Gamma/m)^2$ is second order in $\Gamma$.


\subsection{Spectral representation and sum rule}

Knowing the analytic structure of the dressed propagator above the multi-particle threshold in the first Riemann sheet, namely a pair of complex conjugate poles at $m^2 \pm i m\Gamma$ and a branch cut starting at $4\mu^2$, we can derive the spectral representation of the ghost propagator for $m>2\mu$. 

\begin{figure}[t!]
	\centering
	\subfloat[Subfigure 1 list of figures text][]{
		\includegraphics[scale=0.32]{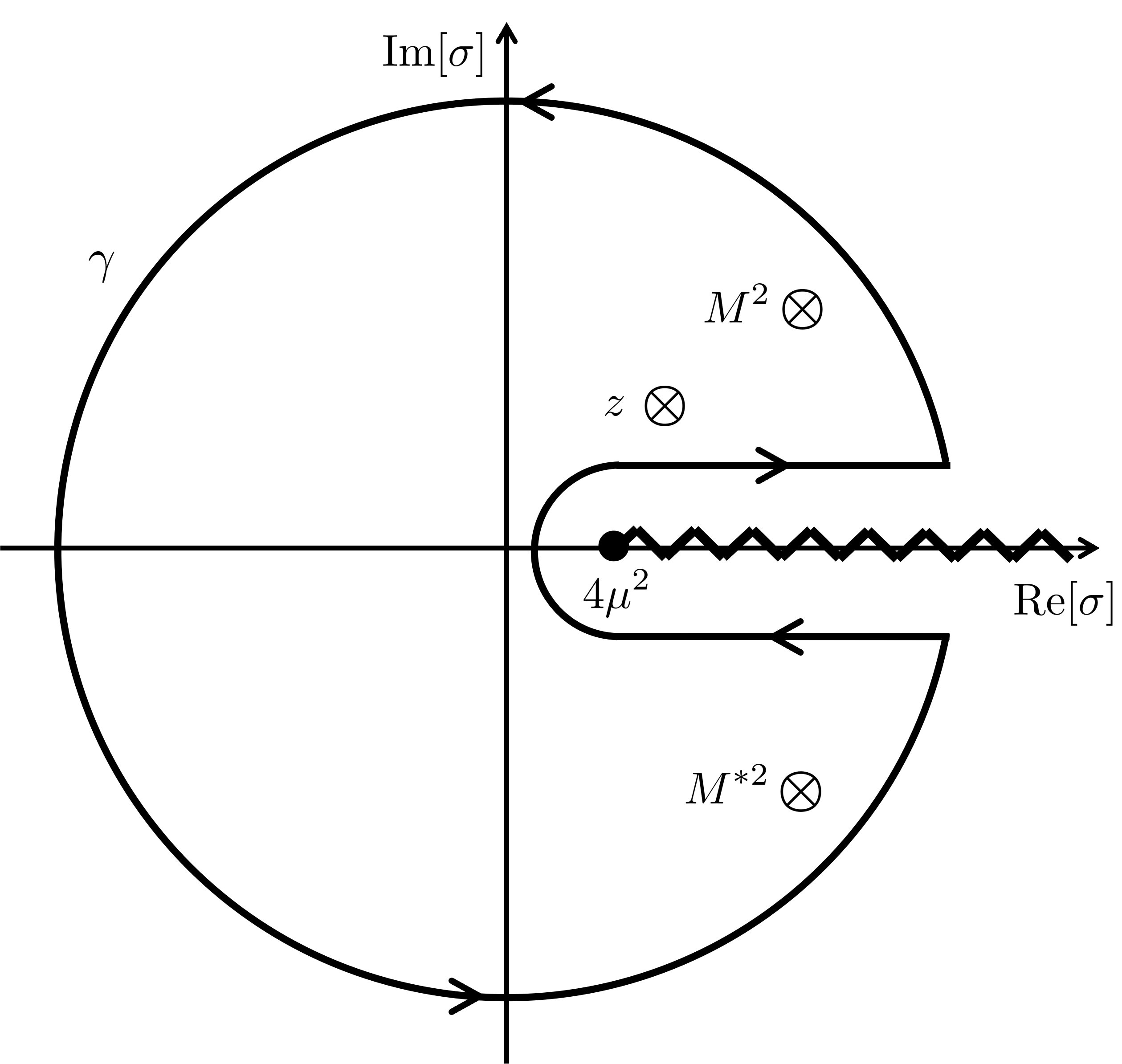}\label{fig2.1}}\qquad\qquad\,
	\subfloat[Subfigure 2 list of figures text][]{
		\includegraphics[scale=0.32]{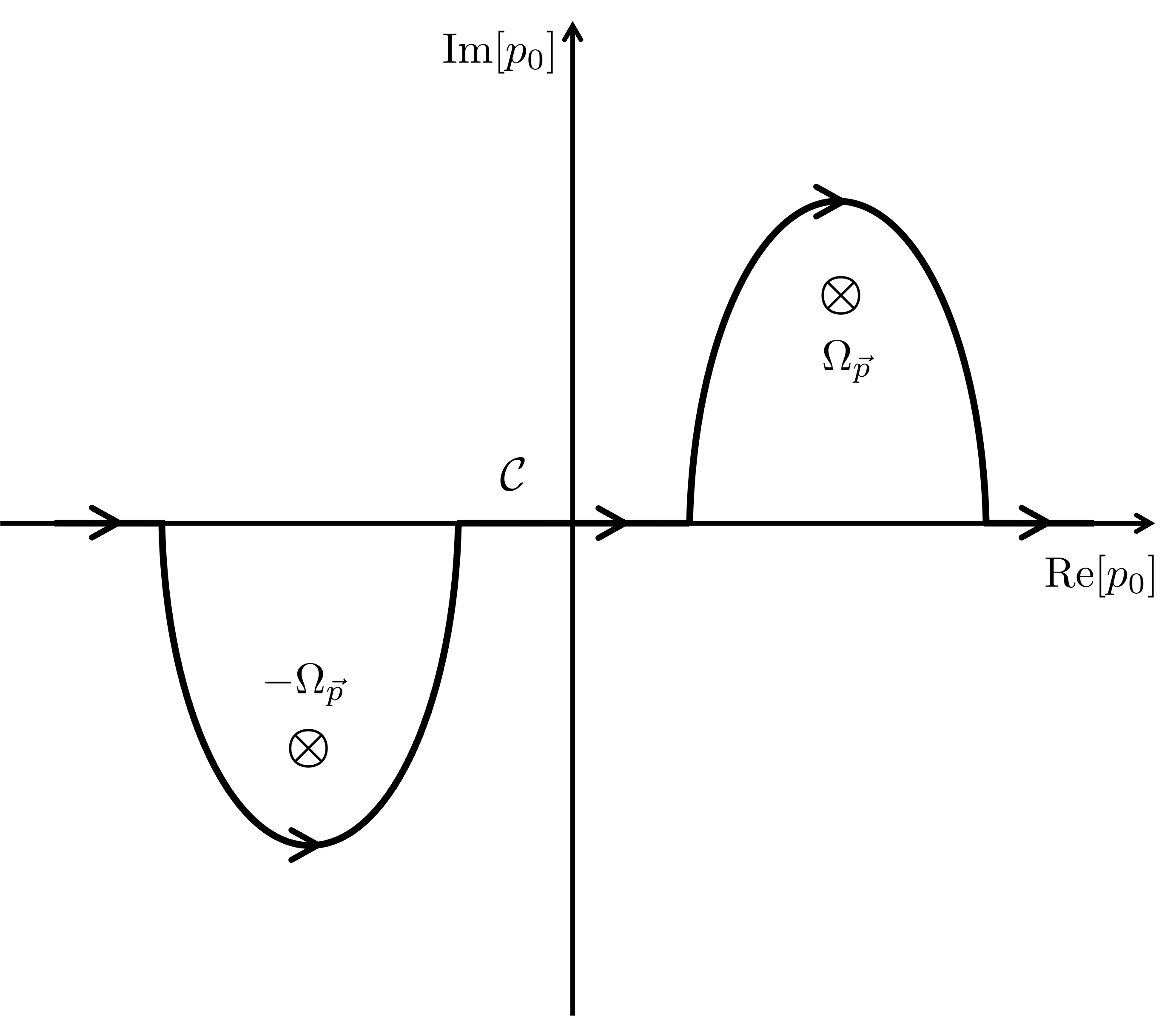}\label{fig2.2}}
	\protect\caption{(a) Integration contour $\gamma$ used in~\eqref{formula-residue-discont} to derive the spectral representation of the ghost propagator~\eqref{spectral-represent}. (b) Lee-Wick contour $\mathcal{C}$ introduced in~\eqref{definition-propags}. Encircled crosses indicate poles, and we recall that $M^2=m^2+im\Gamma$ and $\Omega_{\vec{p}}=\sqrt{\vec{p}^2+M^2}$.}
	\label{fig2}
\end{figure}

Using the relations
\begin{eqnarray}
	\frac{1}{2\pi i}\int_{\gamma}{\rm d}\sigma \frac{\bar{G}_\phi(\sigma)}{\sigma-z}=\frac{1}{2\pi i} \lim\limits_{\varepsilon\rightarrow 0^+}\int_{4\mu^2}^{\infty}{\rm d}\sigma\left[\bar{G}_\phi(\sigma+i\varepsilon)-\bar{G}_\phi(\sigma-i\varepsilon)\right]\frac{1}{\sigma-z}=\int_{4\mu^2}^\infty{\rm d}\sigma\frac{-i\rho(\sigma)}{z-\sigma}\,,
	\label{formula-residue-discont}
\end{eqnarray}
where the integration contour $\gamma$ is shown in Fig.~\ref{fig2.1}, and applying the residue theorem, we obtain the following expression for the dressed propagator as a function of physical momentum~\cite{Coleman:1969xz,Kubo:2024ysu,Buoninfante:2025klm}:
\begin{equation}
	\bar{G}_\phi(-p^2)=\frac{iZ}{p^2+M^2}+\frac{iZ^*}{p^2+M^{*2}}+\int_{4\mu^2}^\infty{\rm d}\sigma \rho(\sigma)\frac{-i}{p^2+\sigma-i\epsilon}\,,
	\label{spectral-represent}
\end{equation}
where $\rho(-p^2)={\rm Im}\big[i\bar{G}_\phi(-p^2+i\epsilon)\big]/\pi>0$, $\epsilon\rightarrow 0^+,$ is the spectral density arising from the discontinuity across the branch cut, whose positivity follows from ${\rm Im}[\Sigma(-p^2+i\epsilon)]>0$ (see~\eqref{imag-self-en}).


The spectral representation in position space is expressed in terms of the expectation value of the time-ordered product in the interacting vacuum $|\bar{0} \rangle$ and of Feynman-like propagators:
\begin{equation}
\left\langle\bar{0}|{\rm T}[\phi(x)\phi(y)]|\bar{0} \right\rangle =-Z\Delta_{\rm F}(x-y;M^2)-Z^*\Delta_{\rm F}(x-y;M^{*2})+\int_{4\mu^2}^\infty{\rm d}\sigma \rho(\sigma)\Delta_{\rm F}(x-y;\sigma)\,,
\label{spectral-represent-position}
\end{equation}
where
\begin{equation}
\Delta_{\rm F}(x-y;s)=\int_{\mathcal{C}} \frac{{\rm d}p_0}{2\pi}\int\frac{{\rm d}^3p}{(2\pi)^3} {\rm e}^{ip\cdot (x-y)} \frac{-i}{p^2+s-i\epsilon}\,,\qquad s=M^2, M^{*2}, \sigma\,.
	\label{definition-propags}
\end{equation}
The integration contour $\mathcal{C}$ is shown in Fig.~\ref{fig2.2} and is known as the Lee-Wick contour~\cite{Lee:1969fy,Lee:1970iw,Lee:chicago}. It coincides with the standard causal Feynman contour for $s=\sigma, M^{*2}$, since it can be deformed to the real line $\mathbb{R}$ without crossing the poles $\pm (\sqrt{\vec{p}^2+\sigma}-i\epsilon)$ and $\pm \Omega^*_{\vec{p}}=\pm \sqrt{\vec{p}^2+M^{*2}}$ which lie in the second and fourth quadrants of the complex $p_0$ plane. In contrast, for $s=M^2$ the contour $\mathcal{C}$ cannot be deformed to the real line due to the presence of the complex poles $\pm \Omega_{\vec{p}}=\pm \sqrt{\vec{p}^2+M^2}$ in the first and third quadrants. It is important to remark that the Lee-Wick contour is not chosen ad hoc but is derived within QFT as a consistent deformation of the causal Feynman contour when the real poles $\pm (\sqrt{\vec{p}^2+m^2}-i\epsilon)$ split into the complex conjugate pairs $\pm (\Omega_{\vec{p}}\,,  \Omega^*_{\vec{p}})$~\cite{Nakanishi:1972wx}.\footnote{The contour deformation can be tracked continuously by studying the ghost mass as a function of the interaction coupling, i.e. $m(g)$. Indeed, in the limit $g\to 0$ we have $\Gamma\to 0^+$, $Z\to 1$, and $\rho(\sigma)\to \delta(\sigma-m^2)$, so that \eqref{spectral-represent} reduces to the free ghost propagator $i/(p^2+m^2-i\epsilon)$, with $m=m(g=0)$.}

Furthermore, using $\left\langle\bar{0}|[\phi(x),\phi(y)]|\bar{0} \right\rangle=\left\langle\bar{0}|{\rm T}[\phi(x)\phi(y)]|\bar{0} \right\rangle-(\left\langle\bar{0}|{\rm T}[\phi(x)\phi(y)]|\bar{0} \right\rangle)^*$, valid for $x^0\geq y^0,$ we can also derive the spectral representation of the commutator~\cite{Kubo:2024ysu}:
\begin{equation}
	\left\langle\bar{0}|[\phi(x),\phi(y)]|\bar{0} \right\rangle =-Z\Delta(x-y;M^2)-Z^*\Delta(x-y;M^{*2})+\int_{4\mu^2}^\infty{\rm d}\sigma \rho(\sigma)\Delta(x-y;\sigma)\,,
	\label{spectral-represent-commutator}
\end{equation}
where 
\begin{equation}
	\Delta(x-y;s)= i\int\frac{{\rm d}^3p}{(2\pi)^3} \frac{\sin\left(\vec{p}\cdot (\vec{x}-\vec{y})-\sqrt{\vec{p}^2+s}\, (x^0-y^0)\right)}{\sqrt{\vec{p}^2+s}}\,,\qquad s=M^2, M^{*2}, \sigma\,,
	\label{definition-Pauli-Jordan}
\end{equation}
is the Pauli-Jordan function generalized to include complex masses.

Using the canonical commutation relation for the ghost field, i.e. $\partial_{x^0}[\left\langle\bar{0}|{\rm T}[\phi(x)\phi(y)]|\bar{0} \right\rangle]_{x^0=y^0}=i\delta^{(3)}(\vec{x}-\vec{y}),$ and the property of the Pauli-Jordan function $\partial_{x^0}[\Delta(x-y;s)]_{x^0=y^0}=-i\delta^{(3)}({\vec{x}-\vec{y}}),$ we obtain the following sum rule from the $x^0$-derivative of~\eqref{spectral-represent-commutator} evaluated at $x^0=y^0$~\cite{Kubo:2024ysu}:  
\begin{eqnarray}
	Z+Z^*=1 + C\,,
	\label{anti-instability-relation}
\end{eqnarray}
where we have defined $C\equiv \int_{4\mu^2}^\infty{\rm d}\sigma \rho(\sigma) >0$.

The quantity $Z+Z^*=2{\rm Re}[Z]$ equals the sum of the residues of the two complex conjugate poles (up to a factor of $i$) and can be interpreted as the probability of having the one-particle ghost state in the asymptotic spectrum. The quantity $C$, instead, measures the probability of finding ordinary multi-particle states. As $C$ increases, so does ${\rm Re}[Z]$, implying that the more likely the ``decay products'' are to be produced, the larger the probability of having a ghost becomes.\footnote{The physical reason a one-particle ghost state cannot decay is unitarity: norm conservation forbids a negative-norm state from evolving into a positive-norm one~\cite{Kubo:2023lpz}. Approaches that allow ghost decay, such as those based on modified diagrammatics~\cite{Donoghue:2019fcb} or alternative inner products~\cite{Salvio:2020axm}, depart from the operator formalism of local QFT.\label{nodecay-unit}} This phenomenon was named \textit{anti-instability} in~\cite{Kubo:2024ysu} and is very different from what happens for an ordinary unstable particle. In the latter case, the relation~\eqref{anti-instability-relation} is replaced by $1=C$, indicating that the unstable particle has decayed and no longer belongs to the asymptotic spectrum, consistently with the absence of first-sheet poles above the multi-particle threshold.

\paragraph{Two questions.} The propagator poles located in the first Riemann sheet are typically associated with the masses of the asymptotic fields and, consequently, with the one-particle asymptotic states, defined in the asymptotic regions $x^0\to\pm\infty$. Ordinary asymptotic fields satisfy the free field equations, and the remnant of the quantum dynamics is encoded in wave-function renormalization constant and renormalized mass. This naturally raises the following main questions: 
\begin{enumerate}

	\item Does the relation~\eqref{anti-instability-relation} imply that there exists a truly free asymptotic ghost field?

	\item Does a ghost field lead to observable negative probabilities and/or complex energies?

\end{enumerate}
The aim of this paper is to provide evidence that the answers to both questions are negative, owing to the nontrivial asymptotic quantum dynamics of the ghost field.

\section{Asymptotic fields} \label{sec:asympt-fields}

To understand the asymptotic behavior of the ghost field, we must identify the real scalar field $\phi^{\rm as}(x)$ to which the Heisenberg field $\phi(x)$ tends, in a weak sense, at asymptotic times, namely
\begin{equation}
	\phi(x)\xrightarrow[x^0\rightarrow \pm \infty]{} Z_\phi^{1/2} \phi^{\rm as}(x)\,,
	\label{phi-asymp}
\end{equation}
and determine its wave-function renormalization constant $Z_\phi\in \mathbb{R}$, along with its Lagrangian, field equation, and propagator, which should reproduce the pole structure in~\eqref{spectral-represent}. 

Two apparent obstacles arise. First, the complex frequency $\Omega_{\vec{p}}$ leads to both exponentially damped and growing contributions at large times that may obscure the meaning of the asymptotic limit~\eqref{phi-asymp}. Indeed, rewriting the position-space propagator~\eqref{spectral-represent-position} as 
\begin{eqnarray}
\!\!\!\!\! \!\!\!\!\!\!\!	\left\langle\bar{0}|{\rm T}[\phi(x)\phi(0)]|\bar{0} \right\rangle \!\!\!\!&=&\!\!\!\! -\int \frac{{\rm d}^3p}{(2\pi)^3}\frac{{\rm e}^{i\vec{p}\cdot \vec{x}}}{2}\left[\theta(x^0) {\rm e}^{-i {\rm Re}[\Omega_{\vec{p}}]x^0}\!\!\left(\frac{Z}{\Omega_{\vec{p}}}{\rm e}^{{\rm Im}[\Omega_{\vec{p}}]x^0}\!+\!\frac{Z^*}{\Omega^*_{\vec{p}}}{\rm e}^{-{\rm Im}[\Omega_{\vec{p}}]x^0}\!\right)\right.\nonumber\\
	\!\!\!\!&&\!\!\!\!\left.+\theta(-x^0) {\rm e}^{i {\rm Re}[\Omega_{\vec{p}}]x^0}\!\!\left(\frac{Z}{\Omega_{\vec{p}}}{\rm e}^{-{\rm Im}[\Omega_{\vec{p}}]x^0}\!+\!\frac{Z^*}{\Omega^*_{\vec{p}}}{\rm e}^{{\rm Im}[\Omega_{\vec{p}}]x^0}\!\right)\!\right]\!+\!\int_{4\mu^2}^\infty\!{\rm d}\sigma \rho(\sigma)\Delta_{\rm F}(x;\sigma)\,,
	\label{spectral-represent-time-dep}
\end{eqnarray}
where we have set $y=0$ for simplicity, we can see explicitly that $(p^2+M^2)^{-1}$ and $(p^2+M^{*2})^{-1}$ generate growing and damped exponentials, respectively, as $x^0 \rightarrow \pm \infty$, complicating the identification of truly free asymptotic one-particle states. Second, there exists no local Hermitian Lagrangian for a single scalar field $\phi^{\rm as}(x)$ whose propagator reproduces the pole structure in~\eqref{spectral-represent}.

In the case of an ordinary field with real frequencies, the asymptotic limit of the multi-particle component of the spectral representation can be shown to be power-law suppressed using the Riemann-Lebesgue lemma; for example, for a two-particle threshold it behaves as $(\mu x^0)^{-3/2}$~\cite{Brown:1992db}. This implies that, if a pole is present, it dominates at very large times\footnote{It is worth recalling that the asymptotic limit $x^0 \rightarrow \pm \infty$ should be physically understood as $|x^0|$ much larger than some characteristic inverse mass scale in the theory, e.g $|x^0| \gg 1/\mu$. The strict limit $x^0 = \pm \infty$ is ill-defined due to the oscillatory terms in the stable pole~\cite{Brown:1992db}. Furthermore, if growing exponentials are present due to complex frequencies, they should be understood in a distributional sense in terms of a complex delta distribution~\cite{Kubo:2023lpz}.}, allowing an unambiguous identification of the asymptotic field and the associated stable one-particle state. Moreover, the asymptotic field satisfies the free field equation with renormalized mass.

When the scalar field is a ghost, the multi-particle contribution exhibits the same power-law decrease, but is not always fully suppressed relative to the pole terms due to the presence of both growing and damped exponentials. Inspecting the time dependence in~\eqref{spectral-represent-time-dep}, one can distinguish at least two different regimes: (\textit{i}) for sufficiently large times but such that $|x^0| < {\rm Im}[\Omega_{\vec{p}}]^{-1}$, the continuum is suppressed relative to both pole terms; (\textit{ii}) for $|x^0| > {\rm Im}[\Omega_{\vec{p}}]^{-1}$, the propagator is dominated by the growing exponentials from the pole term $1/(p^2+M^{2})$, while the continuum component, though power-law suppressed, is larger than the exponentially damped contribution from $1/(p^2+M^{*2})$. These behaviors suggest that free one-particle states can be approximately identified only in the former time regime, while in the latter quantum interference with the multi-particle component could obstruct the existence of free ghost particles asymptotically. 

The physics underlying these unusual features can be understood by \textit{doubling} the asymptotic ghost field, for example through the introduction of a pair of Hermitian conjugate scalar fields $\psi(x)$ and $\psi^\dagger(x)$ such that~\cite{Kubo:2024ysu}
\begin{equation}
	\phi(x)\xrightarrow[x^0\rightarrow \pm \infty]{} Z_\phi^{1/2}\phi^{\rm as}(x)=Z^{1/2}\psi(x)+Z^{*1/2}\psi^\dagger(x)\,,
	\label{psipsi-asymp}
\end{equation}
where $Z\in \mathbb{C}$ is the complex wave-function renormalization constant introduced above.

In what follows, we show that the limit~\eqref{psipsi-asymp} admits a consistent local and Hermitian Lagrangian description of the asymptotic ghost field, whose quantum dynamics exhibits nontrivial interactions and interference with the composite field of the multi-particle state. In particular, we demonstrate that the one-particle ghost state is \textit{masked} by the multi-particle component, so that a quasi-free ghost particle could be defined only approximately, for times $|x^0| < {\rm Im}[\Omega_{\vec{p}}]^{-1}$, implying that no truly \textit{free} asymptotic ghost field or associated one-particle state exists. Furthermore, the exponentially growing frequency contributions turn out to be unambiguous and physically harmless: the norms and equal-time mixed inner products of states remain time-independent~and~finite.

\subsection{Complex-field basis}

Let us now introduce and quantize the system of Hermitian conjugate fields $(\psi, \psi^\dagger)$, which will serve our purpose. Consider the following free Lagrangian density~\cite{Lee:chicago,Kubo:2024ysu,Nakanishi:1972wx}:
\begin{equation}
	\mathcal{L}(\psi,\psi^\dagger)=\frac{1}{2}\left[\partial_\mu \psi \partial^\mu \psi+M^2\psi^2+\partial_\mu \psi^\dagger \partial^\mu \psi^\dagger+M^{*2}\psi^{\dagger 2}\right]\,,
	\label{complex-rep-lagran}
\end{equation}
where the complex conjugate squared masses $M^2 = m^2 + i m \Gamma$ and $M^{*2} = m^2 - i m \Gamma$ are those introduced in the previous section. The field equations are
\begin{equation}
	(\Box-M^2)\psi=0\,,\qquad (\Box-M^{*2})\psi^\dagger=0\,.
	\label{EOM-psipsi}
\end{equation}

We can quantize the system by applying the canonical operator formalism of QFT to~\eqref{complex-rep-lagran}. The field decompositions are~\cite{Nakanishi:1972wx}
\begin{eqnarray}
	\psi(x)\!\!\!&=&\!\!\!\int\frac{{\rm d}^3p}{(2\pi)^{3/2}}\frac{1}{\sqrt{2\Omega_{\vec{p}}}}
	\left(\alpha_{\vec{p}}\,{\rm e}^{-i\Omega_{\vec{p}}x^0+i{\vec{p}\cdot\vec{x}}}+\beta^\dagger_{\vec{p}}\,{\rm e}^{i\Omega_{\vec{p}}x^0-i{\vec{p}\cdot\vec{x}}}\right)\,, \\
	\psi^\dagger(x)\!\!\!&=&\!\!\!\int\frac{{\rm d}^3p}{(2\pi)^{3/2}}\frac{1}{\sqrt{2\Omega^*_{\vec{p}}}}
	\left(\alpha^\dagger_{\vec{p}}\,{\rm e}^{i\Omega^*_{\vec{p}}x^0-i{\vec{p}\cdot\vec{x}}}+\beta_{\vec{p}}\,{\rm e}^{-i\Omega^*_{\vec{p}}x^0+i{\vec{p}\cdot\vec{x}}}\right)\,,
	\label{complex-field-decomp}
\end{eqnarray}
where the complex frequency equals the one defined in the previous section, i.e. $\Omega_{\vec{p}}=\sqrt{\vec{p}^2+M^2}.$
Introducing the conjugate momenta to $\psi(x)$ and $\psi^\dagger(x)$, namely
\begin{eqnarray}
	\pi_\psi(x)=\frac{\delta \mathcal{L}(\psi,\psi^\dagger)}{\delta \dot{\psi}}=-\dot{\psi}\,,\qquad \pi_{\psi^\dagger}(x)=\frac{\delta \mathcal{L}(\psi,\psi^\dagger)}{\delta \dot{\psi}^\dagger}=-\dot{\psi}^\dagger\,,
	\label{comlex-conj-momenta}
\end{eqnarray}
and imposing the canonical commutation relations
\begin{equation}
	\left[\psi(x),\pi_\psi(y)\right]_{x^0=y^0}=i\delta^{(3)}(\vec{x}-\vec{y})\,,\qquad \left[\psi^\dagger(x),\pi_{\psi^\dagger}(y)\right]_{x^0=y^0}=i\delta^{(3)}(\vec{x}-\vec{y})\,,
\end{equation}
we can derive the commutation rules for $\alpha_{\vec{p}}$, $\beta_{\vec{p}}$ and their Hermitian conjugates:
\begin{equation}
\left[\alpha_{\vec{p}},\beta^\dagger_{\vec{k}}\right]=\left[\beta_{\vec{p}},\alpha^\dagger_{\vec{k}}\right]=-\delta^{(3)}\big(\vec{p}-\vec{k}\big)\,,
	\label{CCR-alphabeta}
\end{equation}
while the other commutators can be shown to vanish.

The Hamiltonian of the system reads
\begin{equation}
	H=\int {\rm d}^3x \left[\pi_\psi \dot{\psi}+\pi_{\psi^\dagger} \dot{\psi}^\dagger-\mathcal{L}\right]=-\int {\rm d}^3p  \left[\Omega_{\vec{p}}\, \beta^\dagger_{\vec{p}}\,\alpha_{\vec{p}}+\Omega^*_{\vec{p}}\, \alpha^\dagger_{\vec{p}}\,\beta_{\vec{p}}\right]\,,
	\label{hamilt-alphabeta}
\end{equation}
where standard infinite constants have been subtracted.

Using the commutation relations
\begin{equation}
	\Big[H,\alpha_{\vec{p}}\Big]=-\Omega_{\vec{p}}\,\alpha_{\vec{p}}\,,\quad \Big[H,\alpha^\dagger_{\vec{p}}\Big]=\Omega^*_{\vec{p}}\,\alpha^\dagger_{\vec{p}}\,,\quad \Big[H,\beta_{\vec{p}}\Big]=-\Omega^*_{\vec{p}}\,\beta_{\vec{p}}\,,\quad \Big[H,\beta^\dagger_{\vec{p}}\Big]=\Omega_{\vec{p}}\,\beta^\dagger_{\vec{p}}\,,
\end{equation}
we can interpret $\alpha_{\vec{p}}$ and $\beta_{\vec{p}}$ as annihilation operators, while $\alpha^\dagger_{\vec{p}}$ and $\beta^\dagger_{\vec{p}}$ as creation operators. This allows us to  define the asymptotic vacuum state\footnote{Let us clarify that in this work the state $|\bar{0}\rangle$ denotes the interacting vacuum of the full theory, while $|0\rangle$ (without the overbar) always refers to the asymptotic vacuum, i.e. the state annihilated both by the Hamiltonian~\eqref{hamilt-alphabeta} and by the free Hamiltonian of the $\chi$ field.} $|0\rangle$ through the relations
\begin{equation}
	\alpha_{\vec{p}}|0\rangle=0\,,\qquad \beta_{\vec{p}}|0\rangle=0\,.
\end{equation}
Moreover, we can define the ``one-particle'' complex-energy eigenstates
\begin{equation}
	|\alpha_{\vec{p}}\rangle\equiv \alpha^\dagger_{\vec{p}}|0\rangle\,,\qquad |\beta_{\vec{p}}\rangle\equiv \beta^\dagger_{\vec{p}}|0\rangle\,,
	\label{one-particle-states-alphabeta}
\end{equation}
such that 
\begin{equation}
	H|\alpha_{\vec{p}}\rangle=\Omega^*_{\vec{p}}|\alpha_{\vec{p}}\rangle\,,\qquad H|\beta_{\vec{p}}\rangle=\Omega_{\vec{p}}|\beta_{\vec{p}}\rangle\,.
\end{equation}
A general property of Hermitian Hamiltonians with pairs of eigenstates having complex conjugate eigenvalues is that the norms vanish, while the mixed inner products are nonzero:
\begin{equation}
	\langle\alpha_{\vec{p}}|\alpha_{\vec{k}}\rangle=\langle\beta_{\vec{p}}|\beta_{\vec{k}}\rangle=0\,,\qquad \langle\alpha_{\vec{p}}|\beta_{\vec{k}}\rangle=\langle\beta_{\vec{p}}|\alpha_{\vec{k}}\rangle=-\delta^{(3)}\big(\vec{p}-\vec{k}\big)\,.
	\label{zero-norm}
\end{equation}

We now have all the ingredients to compute the free propagators of the two Hermitian conjugate fields, which can be shown to be equal to
\begin{eqnarray}
\!\!\!	\left\langle0|{\rm T}[\psi(x)\psi(y)]|0 \right\rangle \!\!\!&=&\!\!\! \int_{\mathcal{C}} \frac{{\rm d}p_0}{2\pi}\int\frac{{\rm d}^3p}{(2\pi)^3} {\rm e}^{ip\cdot (x-y)} G_\psi(-p^2) \,,\qquad\,\,\,\, G_\psi(-p^2)=\frac{i}{p^2+M^2}\,, \label{propagators-psipsi}\\[1mm]
\!\!\!	\langle0|{\rm T}[\psi^\dagger(x)\psi^\dagger(y)]|0 \rangle \!\!\!&=&\!\!\! \int_{\mathcal{C}} \frac{{\rm d}p_0}{2\pi}\int\frac{{\rm d}^3p}{(2\pi)^3} {\rm e}^{ip\cdot (x-y)} G_{\psi^\dagger}(-p^2) \,,\qquad G_{\psi^\dagger}(-p^2) =\frac{i}{p^2+M^{*2}}\,,
	\label{propagators-psidaggpsidagg}
\end{eqnarray}
while the mixed time-ordered products vanish. We recall that $\mathcal{C}$ is the Lee-Wick contour shown in Fig.~\ref{fig2.2}, which can be deformed to the real line $\mathbb{R}$ for the $\psi^\dagger$ propagator.

It is easy to see that, using the asymptotic relation in~\eqref{psipsi-asymp}, the propagators~\eqref{propagators-psipsi} and~\eqref{propagators-psidaggpsidagg} turn out to be the right ones to reproduce the correct pole structure in~\eqref{spectral-represent} and~\eqref{spectral-represent-position}:
\begin{eqnarray}
	Z_\phi\,\big\langle0\big|{\rm T}\big[\phi^{\rm as}(x)\phi^{\rm as}(y)\big]\big|0 \big\rangle	\!\!\!&=&\!\!\!\big\langle0\big|{\rm T}\big[\big(Z^{1/2}\psi(x)\!+\!Z^{*1/2}\psi^\dagger(x)\big)\big(Z^{1/2}\psi(y)\!+\!Z^{*1/2}\psi^\dagger(y)\big)\big]\big|0 \big\rangle \nonumber\\[1mm]
	\!\!\!&=&\!\!\! \int_{\mathcal{C}} \frac{{\rm d}p_0}{2\pi}\int\frac{{\rm d}^3p}{(2\pi)^3} {\rm e}^{ip\cdot (x-y)}\left[\frac{iZ}{p^2+M^{2}}+\frac{iZ^*}{p^2+M^{*2}}\right]\,.
	\label{asymp-limit-propag-psipsi}
\end{eqnarray}

\paragraph{Remark.} Although $\psi(x)$ and $\psi^\dagger(x)$ are free asymptotic fields, the corresponding ``one-particle'' states have vanishing norm and thus do not admit an interpretation in terms of freely propagating particles (see also Sec.~\ref{sec:asymp-states}). Moreover, we are ultimately interested in the ghost field $\phi^{\rm as}(x)$ in~\eqref{psipsi-asymp}, which is associated with a negative-norm one-particle state and is the field entering the vertices. In fact, it is possible to construct states with negative or positive norm starting from the zero-norm ones. For example, the states $\frac{1}{\sqrt{2}}(\alpha_{\vec{p}}^\dagger+\beta_{\vec{p}}^\dagger)|0\rangle$ and $\frac{i}{\sqrt{2}}(\alpha_{\vec{p}}^\dagger-\beta_{\vec{p}}^\dagger)|0\rangle$ have negative and positive norm, respectively:
\begin{equation}
\frac{1}{2}\big\langle 0\big|\big(\alpha_{\vec{p}}+\beta_{\vec{p}}\big)\big(\alpha_{\vec{k}}^\dagger+\beta_{\vec{k}}^\dagger\big)\big|0\big\rangle = -\delta^{(3)}\big(\vec{p}-\vec{k}\big)\,,\,\,\quad 
\frac{1}{2}\big\langle 0\big|\big(\alpha_{\vec{p}}-\beta_{\vec{p}}\big)\big(\alpha_{\vec{k}}^\dagger-\beta_{\vec{k}}^\dagger\big)\big|0\big\rangle = \delta^{(3)}\big(\vec{p}-\vec{k}\big)\,.
\label{neg-norm-example}
\end{equation}
What are their corresponding asymptotic fields? Do they admit an interpretation as freely propagating particles? Can they serve as external states to be attached to external legs of Feynman diagrams at $x^0=\pm\infty$? To address these intermediate questions, and in preparation for the main ones listed above, it is useful to switch to a real-field basis. This step will set the semi-final stage to elucidate the quantum dynamics of $\phi^{\rm as}(x)$.

\subsection{Real-field basis}

Introducing the two real fields $\eta(x)$ and $\tilde{\eta}(x)$ via the transformation
\begin{equation}
\left\lbrace \begin{array}{l}
\displaystyle\,\, \psi(x)=\displaystyle \frac{\eta(x)+i\tilde{\eta}(x)}{\sqrt{2}}\\[4mm]
\displaystyle \psi^\dagger(x)=\displaystyle \frac{\eta(x)-i\tilde{\eta}(x)}{\sqrt{2}} 
\end{array}\right. \qquad  \Leftrightarrow\qquad \left\lbrace \begin{array}{l}
\displaystyle \eta(x)=\displaystyle \frac{\psi(x)+\psi^\dagger(x)}{\sqrt{2}}\\[4mm]
\displaystyle \tilde{\eta}(x)=\displaystyle -i\frac{\psi(x)-\psi^\dagger(x)}{\sqrt{2}} 
\end{array}\right.\,,
\label{change-field-basis}
\end{equation}
we can switch from the complex to the real representation, whose Lagrangian density is recast into the following form:
\begin{equation}
	\mathcal{L}\left(\eta,\tilde{\eta}\right)=\frac{1}{2}\left(\partial_\mu\eta\partial^\mu\eta+m^2\eta^2\right)-\frac{1}{2}\left(\partial_\mu\tilde{\eta}\partial^\mu\tilde{\eta}+m^2\tilde{\eta}^2\right)-m\Gamma\, \eta \,\tilde{\eta}\,.
	\label{real-rep-lagran}
\end{equation}
Note that the two real fields have opposite-sign kinetic terms: $\eta(x)$ is a ghost, while $\tilde{\eta}(x)$ is an ordinary field. The corresponding field equations are
\begin{equation}
	(\Box-m^2)\eta(x)=-m\Gamma \tilde{\eta}(x)\,,\qquad (\Box-m^2)\tilde{\eta}(x)=m\Gamma\eta(x)\,.
	\label{real-rep-EOM}
\end{equation}
There is no way to diagonalize~\eqref{real-rep-lagran} in terms of real fields with real masses; indeed, diagonalization brings us back to~\eqref{complex-rep-lagran}. Therefore, $\eta(x)$ and $\tilde{\eta}(x)$ can be viewed as two interacting fields with mass squared $m^2={\rm Re}[M^2]$ and interaction coupling $m\Gamma={\rm Im}[M^2]$, contributing to the asymptotic dynamics of the ghost field.

The conjugate momenta to $\eta(x)$ and $\tilde{\eta}(x)$ are
\begin{eqnarray}
	\pi_{\eta}(x)=\frac{\delta \mathcal{L}\left(\eta,\tilde{\eta}\right)}{\delta \dot{\eta}}=-\dot{\eta}(x)\,,\qquad \pi_{\tilde{\eta}}(x)=\frac{\delta \mathcal{L}\left(\eta,\tilde{\eta}\right)}{\delta \dot{\tilde{\eta}}}=\dot{\tilde{\eta}}(x)\,.
	\label{real-conj-momenta}
\end{eqnarray}
Applying the transformation~\eqref{change-field-basis} to the fields and their conjugate momenta, and using the properties of $\psi(x)$ and $\psi^\dagger(x)$ discussed above, we can readily quantize the real fields $\eta(x)$ and $\tilde{\eta}(x)$. In particular, the canonical commutation relations are
\begin{equation}
	\big[\eta(x),\pi_{\eta}(y)\big]_{x^0=y^0}=i\delta^{(3)}(\vec{x}-\vec{y})\,,\qquad \big[\tilde{\eta}(x),\pi_{\tilde{\eta}}(y)\big]_{x^0=y^0}=i\delta^{(3)}(\vec{x}-\vec{y})\,,
\end{equation}
while the mixed commutators are zero.

We can define time-dependent one-particle states associated with the two real fields:
\begin{eqnarray}
	\left|\eta\big(\vec{p};x^0\big)\right\rangle\!\!\!&\equiv&\!\!\! \int {\rm d}^3x \sqrt{\frac{2|\Omega_{\vec{p}}|}{(2\pi)^3}} {\rm e}^{i\vec{p}\cdot \vec{x}}\eta(x) \big|0\big\rangle= \frac{1}{\sqrt{2}}\left[{\rm e}^{i\theta_{\vec{p}}/2}\alpha_{\vec{p}}^\dagger(x^0)+{\rm e}^{-i\theta_{\vec{p}}/2}\beta_{\vec{p}}^\dagger(x^0)\right] \big|0\big\rangle  \,,\label{eta-state-full}\\ [1mm]
	\left|\tilde{\eta}\big(\vec{p};x^0\big)\right\rangle \!\!\!&\equiv&\!\!\! \int {\rm d}^3x \sqrt{\frac{2|\Omega_{\vec{p}}|}{(2\pi)^3}} {\rm e}^{i\vec{p}\cdot \vec{x}}\tilde{\eta}(x) \big|0\big\rangle= \frac{i}{\sqrt{2}}\left[{\rm e}^{i\theta_{\vec{p}}/2}\alpha_{\vec{p}}^\dagger(x^0)-{\rm e}^{-i\theta_{\vec{p}}/2}\beta_{\vec{p}}^\dagger(x^0)\right] \big|0\big\rangle  \,,
	\label{etatidle-state-full}
\end{eqnarray}
where we have introduced the time-dependent operators $\alpha_{\vec{p}}^\dagger(x^0)\equiv {\rm e}^{i\Omega^*_{\vec{p}}x^0}\alpha^\dagger_{\vec{p}}$ and $\beta^\dagger_{\vec{p}}(x^0)\equiv {\rm e}^{i\Omega_{\vec{p}}x^0}\beta^\dagger_{\vec{p}}$, and used the polar form of the complex frequency $\Omega_{\vec{p}}=|\Omega_{\vec{p}}|{\rm e}^{i\theta_{\vec{p}}}$, with $\theta_{\vec{p}}=\arctan({\rm Im}[\Omega_{\vec{p}}]/{\rm Re}[\Omega_{\vec{p}}])$. We can easily compute norm and equal-time inner products of~\eqref{eta-state-full} and~\eqref{etatidle-state-full}:
\begin{eqnarray}
	\big\langle \eta\big(\vec{p};x^0\big)\big|\eta\big(\vec{k};x^0\big)\big\rangle= -\delta^{(3)}\big(\vec{p}-\vec{k}\big)\cos \theta_{\vec{p}}\,,\qquad 
	\big\langle \tilde{\eta}\big(\vec{p};x^0\big)\big|\tilde{\eta}\big(\vec{k};x^0\big)\big\rangle = \delta^{(3)}\big(\vec{p}-\vec{k}\big)\cos \theta_{\vec{p}}\,,&&	\label{eta-norm-full}\\[1mm]
\big\langle \eta\big(\vec{p};x^0\big)\big|\tilde{\eta}\big(\vec{k};x^0\big)\big\rangle= 
\big\langle \tilde{\eta}\big(\vec{p};x^0\big)\big|\eta\big(\vec{k};x^0\big)\big\rangle = \delta^{(3)}\big(\vec{p}-\vec{k}\big)\sin \theta_{\vec{p}}\,.\qquad \qquad\quad&&	\label{eta-mixed-innerproduct}
\end{eqnarray}
Since both the real and imaginary parts of $\Omega_{\vec{p}}$ are positive and finite, we have $0\leq \theta_{\vec{p}}<\pi/2$, implying $0<\cos\theta_{\vec{p}}\leq 1$. Therefore, the $\eta$ one-particle state has negative norm, while the $\tilde{\eta}$ one-particle state has positive norm. The two states are not orthogonal due to the interaction~in~\eqref{real-rep-lagran}. At zeroth order in the narrow-width approximation $\Gamma/m\ll 1$, i.e. when the interaction is negligible, we have $\theta_{\vec{p}}\simeq 0$, the states become orthogonal, and~\eqref{eta-state-full} and~\eqref{etatidle-state-full}  reduce to the~example~in~\eqref{neg-norm-example}.

One might be led to think that only $\eta(x)$ carries the information about the ghost nature of the field $\phi(x)$, but this is not correct. Our analysis suggests that in the asymptotic limit the ghost field $\phi^{\rm as}(x)$ is composed not only of the ghost field $\eta(x)$ but also of an ordinary field $\tilde{\eta}(x)$, which is tied to $\eta(x)$ through the interaction coupling $m\Gamma$. In fact, we can express the asymptotic ghost field as the following linear combination:
\begin{eqnarray}
Z_\phi^{1/2} \phi^{\rm as}(x)=\sqrt{2|Z|}\,\big[\cos(\theta_{\rm Z}/2)\,\eta(x)-\sin (\theta_{\rm Z}/2)\,\tilde{\eta}(x)\,\big]\,,
\label{phi-as-real-linear-comb}
\end{eqnarray}
where we have used the polar form of $Z=|Z|{\rm e}^{i\theta_{\rm Z}},$ with $\theta_{\rm Z}=\arctan({\rm Im}[Z]/{\rm Re}[Z]).$ This means that both fields will contribute to the pole structure of the $\phi$ propagator.

The propagator of the pair of fields $(\eta,\tilde{\eta})$ is a $2\times 2$ matrix that can be obtained by inverting the kinetic operator in~\eqref{real-rep-lagran}. Working in momentum space, we obtain
\begin{eqnarray}
\hat{G}_{\left(\eta,\tilde{\eta}\right)}(-p^2)= 
\left(\begin{array}{cc}
G_\eta(-p^2)	& G_{\tilde{\eta}\eta}(-p^2)\\[2mm]
G_{\eta\tilde{\eta}}(-p^2)	& G_{\tilde{\eta}}(-p^2)
\end{array}
\right)\,,
\label{propag-matrix}\,
\end{eqnarray}
where 
\begin{eqnarray}
G_{\eta}(-p^2)=-G_{\tilde{\eta}}(-p^2)\!\!\!&=&\!\!\!\frac{1}{2}\left[\frac{i}{p^2+M^2}+\frac{i}{p^2+M^{*2}}\right]\,,\label{propag-eta}\\[1mm]
G_{\eta \tilde{\eta}}(-p^2)=G_{\tilde{\eta}\eta}(-p^2)\!\!\!&=&\!\!\!\frac{1}{2}\left[\frac{1}{p^2+M^2}-\frac{1}{p^2+M^{*2}}\right]\,.
\label{propagators-eta}\,
\end{eqnarray}
Computing the expectation value of the time-ordered product for the linear combination~\eqref{phi-as-real-linear-comb}, it can be straightforwardly shown that all four components of the propagator matrix~\eqref{propag-matrix} contribute to reproduce the correct pole structure in~\eqref{spectral-represent}. Not only are both $G_\eta$ and $G_{\tilde{\eta}}$ relevant, but so are the mixed components $G_{\eta \tilde{\eta}}$ and $G_{\tilde{\eta}\eta}$, due to interference effects induced by the~interaction~in~\eqref{real-rep-lagran}.

As a consistency check of our analysis we note that in the limit $\Gamma\to 0^+$, which also implies ${\rm Im}[Z^{1/2}]=|Z|^{1/2}\sin(\theta_Z/2) \to 0$, the two fields decouple, only the ghost field $\eta(x)$ contributes in~\eqref{phi-as-real-linear-comb}, and the $\eta$ propagator in~\eqref{propag-eta} reduces to the free $\phi$ propagator with real mass.

In summary, above the multi-particle threshold, the asymptotic behavior of the ghost field $\phi(x)$ can be described in terms of another ghost field $\eta(x)$ and an ordinary field $\tilde{\eta}(x)$, which are never free due to the interaction term in~\eqref{real-rep-lagran} controlled by $\Gamma \neq 0$. Our ultimate goal, however, is to determine the asymptotic dynamics of $\phi(x)$ itself -- namely, whether $\phi^{\rm as}(x)$ is a genuinely \textit{free} asymptotic field, and thus whether it is associated with a \textit{free} one-particle state carrying observable negative probabilities and/or complex energies. We now have all the ingredients to address these questions.

\subsection{No free asymptotic ghost field}

The action of $\Box - m^2$ on both sides of~\eqref{psipsi-asymp} or~\eqref{phi-as-real-linear-comb}, using the field equations~\eqref{EOM-psipsi} or~\eqref{real-rep-EOM}, gives
\begin{equation}
(\Box-m^2)Z_\phi^{\rm 1/2}\phi^{\rm as}(x)= -m\Gamma \sqrt{2|Z|}\,\big[\sin(\theta_{\rm Z}/2)\,\eta(x)+\cos (\theta_{\rm Z}/2)\,\tilde{\eta}(x)\,\big] \neq 0\,,
\label{phi-eom}
\end{equation}
which implies that the asymptotic ghost field $\phi^{\rm as}(x)$ is not free! This fact suggests that we can introduce an additional scalar field $\tilde{\phi}^{\rm as}(x)$ with its own wave-function renormalization constant $Z_{\tilde{\phi}}^{\rm 1/2}$ via the definition
\begin{equation}
Z_{\tilde{\phi}}^{\rm 1/2}\tilde{\phi}^{\rm as}(x)\equiv \sqrt{2|Z|}\,\big[\sin(\theta_{\rm Z}/2)\,\eta(x)+\cos (\theta_{\rm Z}/2)\,\tilde{\eta}(x)\,\big]\,,
\label{phitilde-linear-comb}
\end{equation}
so that~\eqref{phi-eom} can be recast as
\begin{equation}
(\Box-m^2)Z_\phi^{\rm 1/2}\phi^{\rm as}(x)= -m\Gamma Z_{\tilde{\phi}}^{\rm 1/2}\tilde{\phi}^{\rm as}(x)\,.
\end{equation}

Inspecting the expressions~\eqref{phi-as-real-linear-comb} and~\eqref{phitilde-linear-comb}, we see that they correspond to a change of field basis via a scaled rotation by an angle $\theta_{\rm Z}/2$. Thanks to this property, we can derive the Lagrangian for $\phi^{\rm as}(x)$, and so for $\tilde{\phi}^{\rm as}(x)$, and analyze their quantum dynamics. Indeed, inverting~\eqref{phi-as-real-linear-comb} and~\eqref{phitilde-linear-comb} we obtain the transformation
\begin{equation}
	\left\lbrace \begin{array}{l}
		\displaystyle \eta(x)=\displaystyle \frac{1}{\sqrt{2|Z|}}\left[Z_\phi^{\rm 1/2}\phi^{\rm as}(x)\cos(\theta_{\rm Z}/2) + Z_{\tilde{\phi}}^{\rm 1/2}\tilde{\phi}^{\rm as}(x) \sin (\theta_{\rm Z}/2)    \right]\\[4.5mm]
		\displaystyle \tilde{\eta}(x)=\displaystyle \frac{1}{\sqrt{2|Z|}}\left[-Z_\phi^{\rm 1/2}\phi^{\rm as}(x)\sin(\theta_{\rm Z}/2) + Z_{\tilde{\phi}}^{\rm 1/2}\tilde{\phi}^{\rm as}(x) \cos (\theta_{\rm Z}/2)   \right]
	\end{array}\right. \,,
	\label{ratation-transf}
\end{equation}
through which the Lagrangian $\mathcal{L}(\eta,\tilde{\eta})$ in~\eqref{real-rep-lagran} transforms into 
\begin{eqnarray}
	\mathcal{L}\left(\phi^{\rm as},\tilde{\phi}^{\rm as}\right)\!\!\!&=&\!\!\!\frac{1}{2}\Big(\partial_\mu\phi^{\rm as}\partial^\mu\phi^{\rm as}+m^2\phi^{{\rm as}\,2}\Big)-\frac{1}{2}\Big(\partial_\mu\tilde{\phi}^{\rm as}\partial^\mu\tilde{\phi}^{\rm as}+m^2\tilde{\phi}^{{\rm as}\,2}\Big)-m\Gamma\, \phi^{\rm as} \,\tilde{\phi}^{\rm as}\nonumber \\[1mm]
	&&\!\!\!+ \tan \theta_{\rm Z}\left[\partial_\mu \phi^{\rm as}\partial^\mu \tilde{\phi}^{\rm as}+m^2\,\phi^{\rm as} \,\tilde{\phi}^{\rm as}+\frac{m\Gamma}{2}\left(\phi^{{\rm as}\,2}-\tilde{\phi}^{{\rm as}\,2}\right) \right]\,,
	\label{phiphitilde-lagran}
\end{eqnarray}
where $\tan \theta_{\rm Z}={\rm Im}[Z]/{\rm Re}[Z].$ We have identified
\begin{equation}
Z_\phi=Z_{\tilde{\phi}}=\frac{2|Z|}{\cos \theta_{\rm Z}}\,,
\label{wavefunction-renorm-form}
\end{equation}
since this is the consistent choice that ensures canonical normalization in the first line of~\eqref{phiphitilde-lagran}. Note that the two kinetic terms have opposite signs; thus, $\tilde{\phi}^{\rm as}(x)$ is an ordinary field.

The field equations are
\begin{eqnarray}
	(\Box-m^2)\phi^{\rm as}(x)\!\!\!&=&\!\!\!-m\Gamma \tilde{\phi}^{\rm as}(x) - \tan \theta_{\rm Z}\left[(\Box-m^2)\tilde{\phi}^{\rm as}(x)-m\Gamma \phi^{\rm as}(x) \right]\,,\\ 
	(\Box-m^2)\tilde{\phi}^{\rm as}(x) \!\!\!&=&\!\!\! m\Gamma\phi^{\rm as}(x)+\tan \theta_{\rm Z}\left[(\Box-m^2)\phi^{\rm as}(x)+m\Gamma \tilde{\phi}^{\rm as}(x) \right]\,,
	\label{real-rep-EOM-phi-as}
\end{eqnarray}
which are solved by the following equivalent system of equations:
\begin{eqnarray}
	(\Box-m^2)\phi^{\rm as}(x)=-m\Gamma \tilde{\phi}^{\rm as}(x)\,, \qquad 
	(\Box-m^2)\tilde{\phi}^{\rm as}(x) = m\Gamma\phi^{\rm as}(x)\,.
	\label{real-rep-EOM-phi-as-equiv}
\end{eqnarray}

The conjugate momenta to $\phi^{\rm as}(x)$ and $\tilde{\phi}^{\rm as}(x)$ are
\begin{eqnarray}
	\pi_{\phi^{\rm as}}(x)\!\!\!&=&\!\!\!\frac{\delta \mathcal{L}\big(\phi^{\rm as},\tilde{\phi}^{\rm as}\big)}{\delta \dot{\phi}^{\rm as}}=-\dot{\phi}^{\rm as}(x)-\tan \theta_{\rm Z}\,\tilde{\phi}^{\rm as}(x)\,,\label{real-conj-momenta-phi-as-1}\\ 
	\pi_{\tilde{\phi}^{\rm as}}(x)\!\!\!&=&\!\!\!\frac{\delta \mathcal{L}\big(\phi^{\rm as},\tilde{\phi}^{\rm as}\big)}{\delta \dot{\tilde{\phi}}^{\rm as}}=\dot{\tilde{\phi}}^{\rm as}(x)-\tan \theta_{\rm Z}\,\phi^{\rm as}(x)\,.
	\label{real-conj-momenta-phi-as-2}
\end{eqnarray}
Carrying over the quantization outlined in the previous subsections to the fields $\phi^{\rm as}(x)$, $\tilde{\phi}^{\rm as}(x)$ and their conjugate momenta, we can canonically quantize~\eqref{phiphitilde-lagran}. In particular, the canonical commutation relations are satisfied:
\begin{equation}
	\Big[\phi^{\rm as}(x),\pi_{\phi^{\rm as}}(y)\Big]_{x^0=y^0}=i\delta^{(3)}(\vec{x}-\vec{y})\,,\qquad \Big[\tilde{\phi}^{\rm as}(x),\pi_{\tilde{\phi}^{\rm as}}(y)\Big]_{x^0=y^0}=i\delta^{(3)}(\vec{x}-\vec{y})\,,
	\label{CCR-phiphitilde}
\end{equation}
and the mixed commutators vanish.

Using the transformations~\eqref{phi-as-real-linear-comb},~\eqref{phitilde-linear-comb}  and the expressions~\eqref{eta-state-full},~\eqref{etatidle-state-full}, we can define time-dependent one-particle states associated with  $\phi^{\rm as}(x)$ and  $\tilde{\phi}^{\rm as}(x)$: 
\begin{eqnarray}
	\big|\phi^{\rm as}\big(\vec{p};x^0\big)\big\rangle\!\!\!&\equiv&\!\!\! \int {\rm d}^3x \sqrt{\frac{2|\Omega_{\vec{p}}|}{(2\pi)^3}} {\rm e}^{i\vec{p}\cdot \vec{x}}\phi^{\rm as}(x) \big|0\big\rangle  \nonumber\\
	\!\!\!&=&\!\!\! \sqrt{\frac{\cos\theta_{\rm Z}}{2}}\left[{\rm e}^{i(\theta_{\vec{p}}-\theta_{\rm Z})/2}\alpha_{\vec{p}}^\dagger(x^0)+{\rm e}^{-i(\theta_{\vec{p}}-\theta_{\rm Z})/2}\beta_{\vec{p}}^\dagger(x^0)\right] \big|0\big\rangle\,, \label{phi-state-full}\\[1.5mm]
	\big|\tilde{\phi}^{\rm as}\big(\vec{p};x^0\big)\big\rangle \!\!\!&\equiv&\!\!\! \int {\rm d}^3x \sqrt{\frac{2|\Omega_{\vec{p}}|}{(2\pi)^3}} {\rm e}^{i\vec{p}\cdot \vec{x}}\tilde{\phi}^{\rm as}(x) \big|0\big\rangle \nonumber\\
	\!\!\!&=&\!\!\! i\sqrt{\frac{\cos\theta_{\rm Z}}{2}}\left[{\rm e}^{i(\theta_{\vec{p}}-\theta_{\rm Z})/2}\alpha_{\vec{p}}^\dagger(x^0)-{\rm e}^{-i(\theta_{\vec{p}}-\theta_{\rm Z})/2}\beta_{\vec{p}}^\dagger(x^0)\right] \big|0\big\rangle \,.
	\label{phitidle-state-full}
\end{eqnarray}
Their norm and equal-time inner products are
\begin{eqnarray}
	\big\langle \phi^{\rm as}\big(\vec{p};x^0\big)\big|\phi^{\rm as}\big(\vec{k};x^0\big)\big\rangle= - \big\langle \tilde{\phi}^{\rm as}\big(\vec{p};x^0\big)\big|\tilde{\phi}^{\rm as}\big(\vec{k};x^0\big)\big\rangle\!\!\!&=&\!\!\! -\delta^{(3)}\big(\vec{p}-\vec{k}\big)\cos (\theta_{\vec{p}}-\theta_{\rm Z})\cos \theta_{\rm Z}\,,	\label{phi-norm-full}\\[1mm]
	\big\langle \phi^{\rm as}\big(\vec{p};x^0\big)\big|\tilde{\phi}^{\rm as}\big(\vec{k};x^0\big)\big\rangle=
	\big\langle \tilde{\phi}^{\rm as}\big(\vec{p};x^0\big)\big|\phi^{\rm as}\big(\vec{k};x^0\big)\big\rangle \!\!\!&=&\!\!\! \delta^{(3)}\big(\vec{p}-\vec{k}\big)\sin (\theta_{\vec{p}}-\theta_{\rm Z})\cos \theta_{\rm Z}\,.	\label{phi-mixed-innerproduct}
\end{eqnarray}
Since both the real and imaginary parts of $\Omega_{\vec{p}}$ and $Z$ are positive and finite (see also the discussion around Fig.~\ref{fig1} on the signs of ${\rm Re}[Z]$ and ${\rm Im}[Z]$), we have $0\leq \theta_{\vec{p}}<\pi/2$ and $0\leq \theta_{\rm Z}<\pi/2$, implying $-\pi/2<\theta_{\vec{p}}-\theta_{\rm Z}<\pi/2$, and thus $0<\cos(\theta_{\vec{p}}-\theta_{\rm Z})\leq 1$. Therefore, the $\phi^{\rm as}$ one-particle state has negative norm, while the $\tilde{\phi}^{\rm as}$ one-particle state has positive norm, as expected. In the limit $\theta_{\rm Z}\to 0$ with $\theta_{\vec{p}}$ kept fixed, we consistently recover the expressions in~\eqref{eta-norm-full} and~\eqref{eta-mixed-innerproduct}.

To confirm that the asymptotic field $\phi^{\rm as}(x)$, obeying the dynamics dictated by the Lagrangian~\eqref{phiphitilde-lagran}, reproduces the pole terms of the dressed ghost propagator in~\eqref{spectral-represent} (up to a factor of $Z_\phi$), it remains to compute the propagator explicitly. Inverting the kinetic operator in~\eqref{phiphitilde-lagran}, we obtain the $2\times 2$ propagator matrix for the pair of fields $(\phi^{\rm as},\tilde{\phi}^{\rm as})$, whose momentum-space~form~is
\begin{eqnarray}
\hat{G}_{\left(\phi^{\rm as},\tilde{\phi}^{\rm as}\right)}(-p^2)= 
\left(\begin{array}{cc}
G_{\phi^{\rm as}}(-p^2) & G_{\tilde{\phi}^{\rm as}\phi^{\rm as}}(-p^2)\\[2.5mm]
G_{\phi^{\rm as}\tilde{\phi}^{\rm as}}(-p^2)	& G_{\tilde{\phi}^{\rm as}}(-p^2)
\end{array}
\right)\,,
\label{propag-matrix-phi}\,
\end{eqnarray}
where 
\begin{eqnarray}
	G_{\phi^{\rm as}}(-p^2)=-G_{\tilde{\phi}^{\rm as}}(-p^2)\!\!\!&=&\!\!\!Z_\phi^{-1}\left[\frac{iZ}{p^2+M^2}+\frac{iZ^*}{p^2+M^{*2}}\right]\,,\label{propag-phi}\\[1mm]
	G_{\phi^{\rm as} \tilde{\phi}^{\rm as}}(-p^2)=G_{\tilde{\phi}^{\rm as}\phi^{\rm as}}(-p^2)\!\!\!&=&\!\!\!Z_\phi^{-1}\left[\frac{Z}{p^2+M^2}-\frac{Z^*}{p^2+M^{*2}}\right]\,,
	\label{propagators-phi-mixed}
\end{eqnarray}
and we recall that $Z_\phi$ is given in~\eqref{wavefunction-renorm-form}. Therefore, the expression for $Z_\phi G_{\phi^{\rm as}}(-p^2)$, or equivalently, the time-ordered product $Z_\phi \langle 0|{\rm T}[\phi^{\rm as}(x)\phi^{\rm as}(y)]|0 \rangle$, reproduces the pole structure of the spectral representation in~\eqref{spectral-represent} and~\eqref{spectral-represent-position}, with the correct  coefficient. 

As a consistency check of our analysis, we note that in the limit $\Gamma\to 0^+$, which also implies $\theta_{\rm Z}\to 0$, and hence $Z_\phi^{-1}Z \to 1/2$ and $Z_\phi^{-1}Z^* \to 1/2$, the $\phi^{\rm as}(x)$ propagator in~\eqref{propag-phi} reduces to the free ghost propagator with real mass.

\subsection{Role of the multi-particle component}\label{sec:role-multi-p}

What remains to be explained about the asymptotic field content is the true identity of the additional ordinary field $\tilde{\phi}^{\rm as}(x)$ and its associated positive-norm one-particle state. We will now show that it corresponds to the asymptotic limit of the composite field $\chi^2(x)$, whose associated state is a superposition of multi-particle (i.e. two-particle) states.

Let us consider the field equation for the Heisenberg field $\phi(x)$,
\begin{equation}
(\Box-m^2)\phi(x)=-\frac{g}{2}\chi^2(x)\,, 
\end{equation}
and the relations
\begin{eqnarray}
\!\!\!\!\!\!\!\!\!\!\!\!\!\!\!\!\!\!\!\!\!\!&&(\Box_x-m^2)(\Box_y-m^2)\big\langle\bar{0}\big|{\rm T}\big[\phi(x)\phi(y)\big]\big|\bar{0} \big\rangle = \frac{g^2}{4}\big\langle\bar{0}\big|{\rm T}\big[\chi^2(x)\chi^2(y)\big]\big|\bar{0} \big\rangle-(\Box_x-m^2)i\delta^{(4)}(x-y)\,,	\nonumber\\[1.5mm]
\!\!\!\!\!\!\!\!\!\!\!\!\!\!\!\!\!\!\!\!\!\!&&(\Box_x-m^2)(\Box_y-m^2)\Delta_{\rm F}(x-y;s)= (s-m^2)^2\Delta_{\rm F}(x-y;s)+(\Box_x+s-2m^2)i\delta^{(4)}(x-y)\,,
\end{eqnarray}
where $s=M^2,M^{*2},\sigma$. 
Then, defining the composite field
\begin{eqnarray}
\tilde{\phi}(x)\equiv \frac{g}{2m\Gamma}\chi^2(x)\,,
\label{def-composite-field}
\end{eqnarray}
and acting with $(\Box_x-m^2)(\Box_y-m^2)$ on~\eqref{spectral-represent-position}, we obtain the following spectral representation:
\begin{eqnarray}
	\big\langle\bar{0}\big|{\rm T}\big[\tilde{\phi}(x)\tilde{\phi}(y)\big]\big|\bar{0} \big\rangle \!\!\!&=&\!\!\!Z\Delta_{\rm F}(x-y;M^2)+Z^*\Delta_{\rm F}(x-y;M^{*2})\nonumber \\[1mm]
	&&\!\!\!+\int_{4\mu^2}^\infty{\rm d}\sigma \rho(\sigma)\frac{(\sigma-m^2)^2}{m^2\Gamma^2}\Delta_{\rm F}(x-y;\sigma) + g(x,y)\,,
	\label{spectral-represent-composite-field}
\end{eqnarray}
where the last piece includes the contact term
\begin{eqnarray}
g(x,y)\equiv \frac{1}{m^2\Gamma^2}\left[2{\rm Im}[Z]m\Gamma +\int_{4\mu^2}^\infty{\rm d}\sigma \rho(\sigma)(\sigma-m^2)\right]i\delta^{(4)}(x-y)\,,
\label{contact-terms}
\end{eqnarray}
which is not important for the purpose of our discussion.

In a standard QFT, where $\phi(x)$ is an ordinary field, the expectation value $\langle \bar{0}|{\rm T}[\chi^2(x)\chi^2(y)]|\bar{0}\rangle$, i.e. the $\chi^2$~propagator, has no pole in the first Riemann sheet above the multi-particle threshold. In contrast, if $\phi(x)$ is a ghost, the  $\chi^2$~propagator shares the same pair of complex conjugate poles as the dressed ghost propagator.

Note that the sign of the pole structure in~\eqref{spectral-represent-composite-field} is opposite to that in~\eqref{spectral-represent-position}. This sign difference precisely matches what we found for the propagators of the asymptotic fields $\phi^{\rm as}(x)$ and $\tilde{\phi}^{\rm as}(x)$ in~\eqref{propag-phi}. This crucial feature allows us to identify the latter field as the asymptotic limit of the composite field~\eqref{def-composite-field} associated with the multi-particle state, namely in a weak sense we have
\begin{eqnarray}
	\tilde{\phi}(x)\xrightarrow[x^0\rightarrow \pm \infty]{} Z_{\tilde{\phi}}^{1/2}\tilde{\phi}^{\rm as}(x)\!\!\!&=&\!\!\!-i\big[Z^{1/2}\psi(x)-Z^{*1/2}\psi^\dagger(x)\big]\nonumber \\
	\!\!\!&=&\!\!\!\sqrt{2|Z|}\,\big[\sin(\theta_{\rm Z}/2)\,\eta(x)+\cos (\theta_{\rm Z}/2)\,\tilde{\eta}(x)\,\big]\,,
	\label{psipsi-asymp-composite}
\end{eqnarray}
which consistently coincide with the definition introduced in~\eqref{phitilde-linear-comb}, where $Z_{\tilde{\phi}}$ is given in~\eqref{wavefunction-renorm-form}.

We can explicitly verify that the antisymmetric combination of Hermitian conjugate fields in~\eqref{psipsi-asymp-composite} correctly reproduces the pole structure in~\eqref{spectral-represent-composite-field}:
\begin{eqnarray}
Z_{\tilde{\phi}}\,\big\langle0\big|{\rm T}\big[\tilde{\phi}^{\rm as}(x)\tilde{\phi}^{\rm as}(y)\big]\big|0 \big\rangle	\!\!\!&=&\!\!\!-\big\langle0\big|{\rm T}\big[\big(Z^{1/2}\psi(x)\!-\!Z^{*1/2}\psi^\dagger(x)\big)\big(Z^{1/2}\psi(y)\!-\!Z^{*1/2}\psi^\dagger(y)\big)\big]\big|0 \big\rangle \nonumber\\[1mm]
\!\!\!&=&\!\!\! \int_{\mathcal{C}} \frac{{\rm d}p_0}{2\pi}\int\frac{{\rm d}^3p}{(2\pi)^3} {\rm e}^{ip\cdot (x-y)}\left[\frac{-iZ}{p^2+M^{2}}+\frac{-iZ^*}{p^2+M^{*2}}\right]\nonumber\\[1mm]
\!\!\!&=&\!\!\! Z\Delta_{\rm F}(x-y;M^2)+Z^*\Delta_{\rm F}(x-y;M^{*2})\,.
\label{asymp-limit-propag-psipsi-composite}
\end{eqnarray}

Analogously to what we have done for the asymptotic ghost field, we can ask the following question: which local and Hermitian Lagrangian for the asymptotic field $\tilde{\phi}^{\rm as}(x)$ reproduces the pole structure in~\eqref{spectral-represent-composite-field}? We already know the answer: the field must be doubled by introducing the ghost field~$\phi^{\rm as}(x)$, thus yielding the Lagrangian in~\eqref{phiphitilde-lagran}.

The identification in~\eqref{psipsi-asymp-composite} is further supported by two additional remarks. First, the one-particle state associated with the composite field $\chi^2(x)$ has positive norm, as it corresponds to a superposition of two-particle $\chi$ states; it can therefore be identified with the positive-norm one-particle state $|\tilde{\phi}^{\rm as}(\vec{p};x^0)\rangle$ in~\eqref{phi-state-full} asymptotically.\footnote{A two-particle state of the elementary field $\chi(x)$ is effectively seen as a one-particle state from the perspective of the composite field $\tilde{\phi}(x)=\frac{g}{2m\Gamma}\chi^2(x)$. It is also worth noting that the poles of the $\chi^2$~propagator in~\eqref{spectral-represent-composite-field} do not correspond to a standard  bound state, as they are complex and lie above the multi-particle threshold in the first Riemann sheet. See~\cite{Asorey:2024mkb,Oda:2026ozf,Oda:2026auw} for different discussions on bound states in certain theories with ghosts.} Indeed, we schematically have
\begin{eqnarray}
\big|\tilde{\phi}^{\rm as}(\vec{p};x^0)\big\rangle\propto \int {\rm d}^3x\, {\rm e}^{i\vec{p}\cdot \vec{x}} \chi^2(x) \big|0\big\rangle = \int\! {\rm d}^3p_1 \int\! {\rm d}^3p_2 \,\,\Pi\big(\vec{p},\vec{p}_1,\vec{p}_2\big)\,\, \big|\chi(\vec{p}_1,\vec{p}_2;x^0)\big\rangle\,,
\label{composite-superposition}
\end{eqnarray}
where $|\chi(\vec{p}_1,\vec{p}_2;x^0)\rangle$ is the (positive-norm) two-particle state of the elementary field $\chi(x)$, and $\Pi(\vec{p},\vec{p}_1,\vec{p}_2)$ is a phase-space factor containing kinematic quantities, including the Dirac delta $\delta^{(3)}(\vec{p}-\vec{p}_1-\vec{p}_2)$.
The~interference between the multi-particle state~\eqref{composite-superposition} and the negative-norm one-particle ghost state $|\phi^{\rm as}(\vec{p};x^0)\rangle$ prevents the latter from admitting a free-particle interpretation.

The second observation is that, inverting the definition in~\eqref{def-composite-field}, we can rewrite the interaction term in the Lagrangian~\eqref{lagrangian} as
\begin{equation}
\frac{g}{2}\phi(x)\chi^2(x)=m\Gamma \phi(x)\tilde{\phi}(x)\,,
\end{equation}
which has a quadratic form similar to that of the interaction terms in the asymptotic Lagrangian~\eqref{phiphitilde-lagran}. This is not just a coincidence, but it has a deeper meaning. Indeed, using the method of Lagrange multiplier, the composite field $\tilde{\phi}(x)$ can be introduced as an auxiliary field in~\eqref{lagrangian} and, upon integrating out the elementary field $\chi(x)$, we can obtain an effective nonlocal Lagrangian $\mathcal{L}_{\rm eff}(\phi,\tilde{\phi})$ for $\phi(x)$ and  $\tilde{\phi}(x)$. We would then expect that the quantum dynamics controlled by this effective description would reproduce the one in~\eqref{phiphitilde-lagran} in the asymptotic limit. The explicit construction of the effective action and the proof of this last statement lie beyond the scope of this work and will be addressed elsewhere.

\subsection{Remarks on asymptotic states}\label{sec:asymp-states}

If the Heisenberg field $\phi(x)$ in~\eqref{lagrangian} were ordinary, the set of asymptotic states above the multi-particle threshold would include only the free one-particle state associated with the asymptotic field $\chi^{\rm as}(x)$. The one-particle state associated with $\phi(x)$ would decay and  be projected out of the asymptotic spectrum, consistently with unitarity~\cite{Veltman:1963th}.

In contrast, if $\phi(x)$ is a ghost, the number of asymptotic fields, instead of decreasing by one, effectively increases by one due to an effective doubling of the ghost field (or equivalently of the composite multi-particle field). We end up with three asymptotic fields: $\chi^{\rm as}(x)$, $\phi^{\rm as}(x)$, and $\tilde{\phi}^{\rm as}(x)\propto (\chi^2)^{\rm as}$. While $\chi^{\rm as}(x)$ satisfies the free field equation with squared mass $\mu^2$, the other two fields interact, are never free asymptotically, and satisfy the field equations in~\eqref{real-rep-EOM-phi-as-equiv}. This then raises the question of which are the true free asymptotic states that can be attached to external legs of Feynman diagrams to build scattering amplitudes. Let us make three remarks.

First, we emphasize that $|\phi^{\rm as}(\vec{p};x^0)\rangle$ cannot be such a state. Owing to its non-orthogonality with the multi-particle state, it cannot be isolated asymptotically. This implies that $|\phi^{\rm as}(\vec{p};x^0)\rangle$ does not admit an interpretation as a freely propagating particle and cannot serve as a free asymptotic state to be attached to external legs of Feynman diagrams. In particular, since the multi-particle component is not washed out asymptotically but instead masks the one-particle ghost state, the standard LSZ construction~\cite{Lehmann:1954rq} cannot be applied to the latter.

Second, one might be tempted to think that, upon diagonalizing the Lagrangian~\eqref{phiphitilde-lagran} via the transformation
\begin{equation}
	\left\lbrace \begin{array}{l}
		\displaystyle \phi^{\rm as}(x)=\displaystyle \sqrt{\frac{\cos \theta_{\rm Z}}{2}}\left[{\rm e}^{i\theta_{\rm Z}/2}\psi(x)+{\rm e}^{-i\theta_{\rm Z}/2}\psi^\dagger(x)\right]\\[4mm]
		\displaystyle \tilde{\phi}^{\rm as}(x)=\displaystyle -i\sqrt{\frac{\cos \theta_{\rm Z}}{2}}\left[{\rm e}^{i\theta_{\rm Z}/2}\psi(x)-{\rm e}^{-i\theta_{\rm Z}/2}\psi^\dagger(x)\right]
	\end{array}\right. \,,
	\label{change-field-basis-to-complex}
\end{equation}
the states associated with the Hermitian conjugate fields can then be identified as the correct external asymptotic states. However, although $\psi(x)$ and $\psi^\dagger(x)$ are free asymptotic fields, as they satisfy the free field equations~\eqref{EOM-psipsi}, the corresponding ``one-particle'' states are not suitable candidates for  external LSZ-like states. Indeed, the states
\begin{eqnarray}
\!\!\!\!\!\!\!\!\!\!\!\!\!\!\!\!	\left|\psi\big(\vec{p};x^0\big)\right\rangle\!\!\!&\equiv&\!\!\! \int {\rm d}^3x \sqrt{\frac{2\Omega_{\vec{p}}}{(2\pi)^3}} {\rm e}^{i\vec{p}\cdot \vec{x}}\psi(x) \big|0\big\rangle= \frac{{\rm e}^{i(\theta_{\vec{p}}-\theta_{\rm Z})/2}}{\sqrt{2 \cos\theta_{\rm Z}}}\Big[ \big|\phi^{\rm as}\big(\vec{p};x^0\big)\big\rangle +i \big|\tilde{\phi}^{\rm as}\big(\vec{p};x^0\big)\big\rangle\Big]\,,\label{psi-state-full}\\ [1mm]
\!\!\!\!\!\!\!\!\!\!\!\!\!\!\!\!	\left|\psi^\dagger\big(\vec{p};x^0\big)\right\rangle\!\!\!&\equiv&\!\!\! \int {\rm d}^3x \sqrt{\frac{2\Omega_{\vec{p}}^*}{(2\pi)^3}} {\rm e}^{i\vec{p}\cdot \vec{x}}\psi^\dagger(x) \big|0\big\rangle= \frac{{\rm e}^{-i(\theta_{\vec{p}}-\theta_{\rm Z})/2}}{\sqrt{2\cos\theta_{\rm Z}}}\Big[ \big|\phi^{\rm as}\big(\vec{p};x^0\big)\big\rangle -i \big|\tilde{\phi}^{\rm as}\big(\vec{p};x^0\big)\big\rangle\Big]\,,
\label{psidagger-state-full}
\end{eqnarray}
corresponding to the time-dependent version of the ones in~\eqref{one-particle-states-alphabeta}, are zero-norm superpositions of one-particle and multi-particle states. Therefore, they do not admit a particle interpretation.\footnote{It is instructive to compare with neutrino mixing, where one also has diagonal (mass) and non-diagonal (flavor) bases. Despite this similarity, the situations are fundamentally different. In the neutrino case, diagonalization yields standard free asymptotic one-particle states with positive norm, which can be used as external LSZ states to construct amplitudes; flavor amplitudes are then obtained as certain combinations of those computed in the mass basis~\cite{Giunti:2007ry}. In contrast, in our case diagonalization yields zero-norm states that are superpositions of one-particle and multi-particle states, and thus cannot serve as asymptotic external states in scattering amplitudes.}

Third, one may want to consider an orthogonal basis of states, in the hope that these admit a particle interpretation and can serve as genuine LSZ-like states. An orthogonal basis can be easily obtained via a scaled rotation and consists of the following two states:
\begin{eqnarray}
\!\!\!\!\!\!\!\!\!\!\!\frac{\left|\psi^\dagger\big(\vec{p};x^0\big)\right\rangle \!+\! \left|\psi\big(\vec{p};x^0\big)\right\rangle}{\sqrt{2}} \!\!\!\!&=&\!\!\! \frac{1}{\sqrt{\cos\theta_{\rm Z}}} \left[\cos\!\left(\!\frac{\theta_{\vec{p}}\!-\!\theta_{\rm Z}}{2}\!\right)\! \big|\phi^{\rm as}\big(\vec{p};x^0\big)\big\rangle - \sin\!\left(\!\frac{\theta_{\vec{p}}\!-\!\theta_{\rm Z}}{2}\!\right) \! \big|\tilde{\phi}^{\rm as}\big(\vec{p};x^0\big)\big\rangle\right],\label{state-plus}\\ [1mm]
\!\!\!\!\!\!\!\!\!\!\!i\frac{\left|\psi^\dagger\big(\vec{p};x^0\big)\right\rangle \!-\! \left|\psi\big(\vec{p};x^0\big)\right\rangle}{\sqrt{2}}\!\!\!\!&=&\!\!\! \frac{1}{\sqrt{\cos\theta_{\rm Z}}} \left[\sin\!\left(\!\frac{\theta_{\vec{p}}\!-\!\theta_{\rm Z}}{2}\!\right)\! \big|\phi^{\rm as}\big(\vec{p};x^0\big)\big\rangle + \cos\!\left(\!\frac{\theta_{\vec{p}}\!-\!\theta_{\rm Z}}{2}\!\right) \! \big|\tilde{\phi}^{\rm as}\big(\vec{p};x^0\big)\big\rangle\right],
\label{state-minus}
\end{eqnarray}
which have negative and positive norm, respectively, and correspond to the time-dependent version of the states in~\eqref{neg-norm-example}. Again, we obtained superpositions of one-particle ghost and multi-particle states, preventing~\eqref{state-plus} and~\eqref{state-minus} from being free  one-particle states, despite their orthogonality. Only in the limit of zero interactions $\Gamma \to 0$, namely $\theta_{\vec{p}}, \theta_{\rm Z} \to 0$, the state~\eqref{state-plus} would become a true free asymptotic one-particle state orthogonal to the multi-particle component. This remark further confirms that the interacting nature of the asymptotic ghost field does not allow the existence of a freely propagating ghost particle at asymptotic times.

In summary, the QFT framework applied to the Lagrangian~\eqref{lagrangian} shows that, above the multi-particle threshold, the set of asymptotic fields is effectively given by $\lbrace \chi^{\rm as}(x), \phi^{\rm as}(x), \tilde{\phi}^{\rm as}(x)\rbrace$. However, only $\chi^{\rm as}(x)$ is associated with a true free asymptotic one-particle state admitting a particle interpretation. In other words, scattering amplitudes defined in an infinite interval of time can have only $\chi$ states as external ones. The states $|\phi^{\rm as}(\vec{p};x^0)\rangle$ and $|\tilde{\phi}^{\rm as}(\vec{p};x^0)\rangle$ also exist asymptotically, but they mix and correspond to non-stationary configurations that do not describe freely propagating particles.

The nontrivial interaction and interference between the one-particle ghost state and the multi-particle component agree with the initial intuition outlined at the beginning of Sec.~\ref{sec:asympt-fields}. To better understand the physical implications of the asymptotic quantum dynamics for the ghost field, in the next section we will examine the role of the interactions in~\eqref{phiphitilde-lagran} in more detail.

\section{Quantum dynamical effects}\label{sec:compl-quant-eff}

Inspecting the Lagrangian~\eqref{phiphitilde-lagran}, we identify two types of interaction terms: one proportional to $m\Gamma={\rm Im}[M^2]$ in the first line, and another proportional to $\tan\theta_{\rm Z}={\rm Im}[Z]/{\rm Re}[Z]$ in the second line. These two control complementary quantum dynamical effects.

\subsection{Classical remnant}

The interaction coupling $m\Gamma$ is the only one that effectively carries information about the quantum dynamics to the classical level, since it directly affects the field equations in~\eqref{real-rep-EOM-phi-as-equiv}. Indeed, the interaction terms controlled by $\Gamma\sim\mathcal{O}(\hbar)$ represent pure classical remnants of the quantum dynamics. On the other hand, $Z$ does not enter explicitly in the classical evolution.

\subsection{Multi-particle masking} 

The inverse imaginary part of the complex mass controls how rapidly the non-orthogonality of the negative-norm ghost and positive-norm (multi-particle)  states~\eqref{phi-state-full} and~\eqref{phitidle-state-full} becomes $\mathcal{O}(1)$. This can be understood by examining the square of the unequal-time mixed inner product between the two normalized states:
\begin{equation}
\frac{\big|\big\langle \phi^{\rm as}\big(\vec{p};0\big)\big|\tilde{\phi}^{\rm as}\big(\vec{p};t\big)\big\rangle\big|^2}{\big|\big\langle \phi^{\rm as}\big(\vec{p};0\big)\big|\phi^{\rm as}\big(\vec{p};0\big)\big\rangle\big| \big\langle \tilde{\phi}^{\rm as}\big(\vec{p};t\big)\big|\tilde{\phi}^{\rm as}\big(\vec{p};t\big)\big\rangle}= \frac{1}{\cos^2(\theta_{\vec{p}}-\theta_{\rm Z})}\left[\sinh^2\left({\rm Im}[\Omega_{\vec{p}}]\,t\right)+\sin^2\left(\theta_{\vec{p}}-\theta_{\rm Z}\right)\right]\,,
\label{phi-mixed-innerproduct-squared}
\end{equation}
where the modulus in the denominator arises because the normalized ket (with norm $-1$), i.e. $|\phi^{\rm as}(\vec{p};0)\rangle/\sqrt{\langle \phi^{\rm as}(\vec{p};0)|\phi^{\rm as}(\vec{p};0)\rangle}$,  carries a factor of $i$ in the denominator due to its negative norm.

To gain physical insight, let us work in the narrow-width approximation and consider a regime in which at $t=0$ the two states are approximately orthogonal, and ask how long it takes for the overlap to become $\mathcal{O}(1)$. If $\Gamma/m\ll 1$, we can write
\begin{eqnarray}
&&\!\!\!\!\!\!\!\!\!\!\!\!\!\!\!\!\!\!{\rm Re}[\Omega_{\rm p}]=\sqrt{\frac{\sqrt{(\vec{p}^2+m^2)^2+m^2\Gamma^2}+\vec{p}^2+m^2}{2}}\simeq \sqrt{\vec{p}^2+m^2}+\mathcal{O}(\Gamma^2)\,, \\[1.5mm]
&&\!\!\!\!\!\!\!\!\!\!\!\!\!\!\!\!\!\!{\rm Im}[\Omega_{\rm p}]=\sqrt{\frac{\sqrt{(\vec{p}^2+m^2)^2+m^2\Gamma^2}-\vec{p}^2-m^2}{2}}\simeq \frac{m\Gamma}{2\sqrt{\vec{p}^2+m^2}}+\mathcal{O}(\Gamma^3)\,, \\[1.5mm]
&&\!\!\!\!\!\!\!\!\!\!\!\!\!\!\!\!\!\!M\simeq m+i\frac{\Gamma}{2}+\mathcal{O}(\Gamma^2)\,,\qquad \cos^{-2}(\theta_{\vec{p}}-\theta_{\rm Z}) \simeq 1+\mathcal{O}(\Gamma^2)\,,\qquad
\sin^{2}(\theta_{\vec{p}}-\theta_{\rm Z})\simeq  \mathcal{O}(\Gamma^2) \,,
\label{approxim-relations}
\end{eqnarray}
Then, working at rest (i.e. $\vec{p}=0$) for simplicity, \eqref{phi-mixed-innerproduct-squared} becomes\footnote{The quantities in~\eqref{phi-mixed-innerproduct-squared} and~\eqref{phi-mixed-innerproduct-squared-approx} should not be viewed as probabilities, but as physical measures of non-orthogonality. Their unbounded behavior does not conflict with a probabilistic interpretation.}
\begin{equation}
	\frac{\big|\big\langle \phi^{\rm as}(0;0)\big|\tilde{\phi}^{\rm as}(0;t)\big\rangle\big|^2}{\big|\big\langle \phi^{\rm as}(0;0)\big|\phi^{\rm as}(0;0)\big\rangle \big|\big\langle \tilde{\phi}^{\rm as}(0;t)\big|\tilde{\phi}^{\rm as}(0;t)\big\rangle}\simeq \sinh^2\left(\frac{\Gamma t}{2}\right)+\mathcal{O}(\Gamma^2)\,.
	\label{phi-mixed-innerproduct-squared-approx}
\end{equation}
We see that the inverse of the imaginary part of the complex mass, i.e. ${\rm Im}[M]^{-1} \simeq 2/\Gamma$, sets the timescale after which the non-orthogonality becomes $\mathcal{O}(1)$. In other words, after a time $t \sim 2/\Gamma$, the negative-norm one-particle ghost state becomes \textit{masked} by a superposition of positive-norm multi-particle (i.e. two-particle) states. In a generic boosted frame $(\vec{p} \neq 0)$, the timescale is $(\sqrt{\vec{p}^2 + m^2}/m)(2/\Gamma) \simeq {\rm Im}[\Omega_{\vec{p} \neq 0}]^{-1} > 2/\Gamma$.

The true free asymptotic states are instead those in~\eqref{one-particle-states-alphabeta}: linear combinations of the negative-norm~\eqref{phi-norm-full} and positive-norm~\eqref{phitidle-state-full} states with zero norm and hence vanishing localization probability. This physical scenario can be interpreted as saying that the positive-norm state~\eqref{phitidle-state-full} associated with the additional asymptotic field $\tilde{\phi}^{\rm as}(x)$ \textit{masks} the negative-norm state~\eqref{phi-norm-full}, thus rendering any  potential negative probability associated with the ghost $\phi^{\rm as}(x)$ unobservable.

\subsection{Quantum correlations}

We have shown that the canonical commutation relations~\eqref{CCR-phiphitilde} are satisfied for the interacting conjugate momenta defined in~\eqref{real-conj-momenta-phi-as-1} and~\eqref{real-conj-momenta-phi-as-2}. This implies that the mixed commutators involving the would-be free conjugate momenta $-\dot{\phi}^{\rm as}(x)$ and $\dot{\tilde{\phi}}^{\rm as}(x)$ are nonzero. Indeed, we have
\begin{eqnarray}
	\Big[\phi^{\rm as}(x),-\dot{\phi}^{\rm as}(y)\Big]_{x^0=y^0}=  \Big[\tilde{\phi}^{\rm as}(x),\dot{\tilde{\phi}}^{\rm as}(y)\Big]_{x^0=y^0}\!\!\!&=&\!\!\! i\delta^{(3)}(\vec{x}-\vec{y}) \cos^2\theta_{\rm Z}\,,\\[1mm]
	\Big[\phi^{\rm as}(x),\dot{\tilde{\phi}}^{\rm as}(y)\Big]_{x^0=y^0}=  \Big[\tilde{\phi}^{\rm as}(x),\dot{\phi}^{\rm as}(y)\Big]_{x^0=y^0}\!\!\!&=&\!\!\! -i\delta^{(3)}(\vec{x}-\vec{y}) \sin\theta_{\rm Z}\,\cos\theta_{\rm Z}\,.
	\label{CCR-phiphitilde-entanglement}
\end{eqnarray}
In particular, the nonvanishing commutators in~\eqref{CCR-phiphitilde-entanglement} signal some form of quantum correlations between the two asymptotic fields that would not exist if ${\rm Im}[Z]=0$. This quantum effect is directly tied to the coupling $\tan\theta_{\rm Z}$ and does not depend explicitly on ${\rm Im}[M^2]$. As a consistency check, note that in the limit ${\rm Im}[Z]\to 0$ with $m\Gamma$ kept fixed, we recover the same structure for the Lagrangian, conjugate momenta, and commutation relations of the pair $(\eta,\tilde{\eta})$, as expected. 

\subsection{Quantum interference}

The underlying quantum property driving these complementary dynamical effects, shared by both interaction couplings ${\rm Im}[M^2]=m\Gamma$ and ${\rm Im}[Z]$, is quantum interference. This can be understood from the nonvanishing off-diagonal elements of the propagator matrices~\eqref{propag-matrix-phi} and~\eqref{propag-matrix}, which stem from the quadratic off-diagonal terms in the Lagrangians~\eqref{phiphitilde-lagran} and~\eqref{real-rep-lagran}. It is worth emphasizing that without these interference terms, the correct complex conjugate pole structure of the ghost propagator, as dictated by the spectral representation~\eqref{spectral-represent} or~\eqref{spectral-represent-position}, could not be reproduced.

Furthermore, the non-orthogonality between the negative- and positive-norm states associated with the mixed asymptotic fields $\phi^{\rm as}(x)$ and $\tilde{\phi}^{\rm as}(x)$ may be viewed as a manifestation of quantum interference, indicating that no well-defined notion of energy can be assigned~to~these~states.

\section{Physical meaning of the complex mass} \label{sec:phys-impl}

For an ordinary unstable particle, the real and imaginary parts of the complex mass have a clear physical interpretation: $m$ is the physical mass (rest energy) of the propagating particle prior to decay, while $2/\Gamma$ sets the timescale of its lifetime. Additionally, in the language of resonances, $m$ is the energy peak of the Breit-Wigner function, while $\Gamma$ its width. What, instead, is the interpretation of the complex mass characterizing the asymptotic ghost field?

First of all, we note that if we are interested in the resonant behavior of the dressed ghost propagator, $m$ and $\Gamma$ can still be assigned the meaning of energy peak and width of a Breit-Wigner-like function~\cite{Buoninfante:2025klm}. Moreover, we have also learnt that ${\rm Im}[M^2]=m\Gamma$ has the meaning of interaction coupling between the asymptotic fields $\phi^{\rm as}(x)$ and $\tilde{\phi}^{\rm as}(x)$, i.e. between the asymptotic ghost field and the multi-particle component. 

What about an interpretation in terms of physical mass and timescale?

\subsection{$2/\Gamma$ as a timescale}

In the previous section we found that, in the narrow-width approximation, $2/\Gamma$ is the timescale after which the non-orthogonality between the negative- and positive-norm states associated with $\phi^{\rm as}(x)$ and $\tilde{\phi}^{\rm as}(x)$ becomes $\mathcal{O}(1)$. Let us elaborate more on this point. 

For a narrow ghost resonance, i.e. $\Gamma/m\ll 1$, the interaction couplings between the two asymptotic fields in~\eqref{phiphitilde-lagran} are negligible. Additionally, if $t\ll 2/\Gamma$, the quantity in~\eqref{phi-mixed-innerproduct-squared-approx} is approximately zero, so the two states can be considered almost orthogonal. This means that for times much shorter than $2/\Gamma$ the ghost field can be treated as approximately free, admitting a short-time, approximate free-particle interpretation. Conversely, once $t\gtrsim 2/\Gamma$, the field $\phi^{\rm as}(x)$ starts to appreciably feel the presence of $\tilde{\phi}^{\rm as}(x)$, and quantum interference with the multi-particle component becomes significant, so that no notion of free particle applies, not even approximately. Therefore, $\tau_{\rm m}\equiv 2/\Gamma$ can be interpreted as the rest-frame \textit{masking time} after which the interaction between the one-particle ghost and the multi-particle states becomes significant, with their overlap reaching order one, so that the ghost gets masked by the multi-particle component.

This explanation is consistent with the intuition outlined at the beginning of Sec.~\ref{sec:asympt-fields}, where less rigorous arguments were formulated in terms of the boosted~timescale~${\rm Im}[\Omega_{\vec{p}}]^{-1}$.

\subsection{$m$ as an approximate free-particle mass}

In the narrow-width approximation $\Gamma/m \ll 1$ and for short times $t \ll 2/\Gamma$, $m$ can be interpreted as the physical mass of a freely propagating ghost particle. Indeed, at leading order, the complex frequency reduces to $\Omega_{\vec{p}} \simeq m + \mathcal{O}(\Gamma)$. However, this interpretation breaks down~as~soon~as~$t \sim 2/\Gamma$.

This situation parallels that of an ordinary unstable particle, which behaves approximately as freely propagating only on timescales much shorter than its lifetime, before interactions driving its decay become relevant. The key difference is that, while an unstable particle eventually decays and leaves the asymptotic spectrum, a one-particle ghost state cannot decay (see the discussion around footnote~\ref{nodecay-unit}) and, moreover, interacts nontrivially with the multi-particle component even asymptotically. This prevents the existence of a free asymptotic one-particle ghost state. 

An equivalent way to phrase the above interpretation is that a freely propagating ghost particle is confined to times much shorter than $2/\Gamma$, after which the negative-norm one-particle state becomes masked by a superposition of positive-norm multi-particle states. Consequently, a detector can never observe an isolated ghost particle asymptotically.

\section{Conclusions and outlook} \label{sec:concl-outl}

We can now provide answers to the two questions posed at the end of Sec.~\ref{sec:field-theory-model}.
\begin{enumerate}
	
	\item No, the ghost field is never free asymptotically due to nontrivial interactions with the multi-particle component that persist at asymptotic times, inducing quantum correlations and interference effects. The anti-instability relation~\eqref{anti-instability-relation} only implies that one-particle ghost states cannot decay and thus remain part of the asymptotic spectrum; it does not, however, provide any additional information about the quantum dynamics.
	
	\item No, we do not find evidence for observable negative probabilities and complex energies associated with the ghost field $\phi(x)$. The negative-norm one-particle state strongly overlaps with a superposition of positive-norm multi-particle states associated with the composite field $\tilde{\phi}(x)=\frac{g}{2m\Gamma}\chi^2(x)$. As a result, a ghost gets masked by the multi-particle component and does not admit a particle interpretation at asymptotic times. Moreover, the complex poles in~\eqref{spectral-represent} do not lead to observable complex energies. In fact, the real and (inverse) imaginary parts of the complex mass admit physical interpretations in terms of physical real mass (${\rm Re}[M]\simeq m$), timescale~(${\rm Im}[M]^{-1}\simeq 2/\Gamma$), and interaction coupling (${\rm Im}[M^2]=m\Gamma$).
	
\end{enumerate}

Let us emphasize that our findings rely on the operator formalism of local QFT. This conservative assumption requires that the ghost field $\phi(x)$ in~\eqref{lagrangian} must be perturbatively quantized in an indefinite-norm vector space to ensure unitarity and a bounded Hamiltonian.\footnote{In particular, neither modified diagrammatic rules~\cite{Donoghue:2019fcb,Anselmi:2021hab} nor new inner products or alternative probabilistic interpretations~\cite{Salvio:2015gsi,Strumia:2017dvt,Holdom:2024onr} were needed to perturbatively quantize and make sense of the ghost in our study.} Moreover, a local and Hermitian Lagrangian description at asymptotic times requires an effective doubling of the ghost field to account for the pair of complex conjugate poles in the first Riemann sheet of the dressed propagator. The additional asymptotic field is identified with the asymptotic limit of the composite field associated with a superposition of multi-particle (i.e.~two-particle) states.

If we tried to guess a local and Hermitian Lagrangian for the asymptotic ghost field that reproduced the pole structure of the propagator~\eqref{spectral-represent} with the correct coefficients, it would be unlikely to easily arrive at the nontrivial form in~\eqref{phiphitilde-lagran}. However, by exploiting properties of the complex-field $(\psi, \psi^\dagger)$ and real-field $(\eta, \tilde{\eta})$ bases, we managed to derive~\eqref{phiphitilde-lagran} and identify its interactions. The interaction coupling ${\rm Im}[M^2]=m\Gamma$ mainly controls the classical remnant of the quantum dynamics and the non-orthogonality of the negative-norm and positive-norm states. The interaction coupling $\tan\theta_{\rm Z}={\rm Im}[Z]/{\rm Re}[Z]$ generates nontrivial quantum correlations between the two asymptotic fields. Both interactions are responsible for quantum interference. 

The results of this work provide a novel perspective on QFTs with indefinite-norm ghosts, supporting their physical consistency. However, several open questions must be addressed to make these conclusions definitive. Before concluding, we briefly comment on some of these issues and discuss potential future directions.

\medskip

\noindent \textbf{Scattering amplitudes.} As explained in Sec.~\ref{sec:asymp-states}, the interacting nature of the asymptotic ghost field precludes the existence of LSZ-like one-particle ghost states that can be attached to external legs of Feynman diagrams in the construction of scattering amplitudes between asymptotic times. However, as discussed in Secs.~\ref{sec:compl-quant-eff} and~\ref{sec:phys-impl}, quasi-free one-particle ghost states can be defined for timescales much shorter than the inverse width, i.e. $t \ll 2/\Gamma$. This suggests that alternative or even new methods for constructing scattering states at finite times, such as those proposed in~\cite{Collins:2019ozc,Anselmi:2023phm}, may be required for this type of problem. In particular, the formalism in~\cite{Collins:2019ozc} might be useful for understanding~\eqref{psipsi-asymp} and~\eqref{psipsi-asymp-composite} as strong limits as well, rather than just in a weak sense.

\medskip

\noindent \textbf{Effective-action approach.} We have analyzed in detail the interaction between the one-particle ghost state and the multi-particle state in terms of asymptotic fields. To further understand the underlying quantum dynamics, however, a study at the level of Heisenberg fields is required. In particular, we must clarify how the ghost field $\phi(x)$ and the composite field $\tilde{\phi}(x)$ behave. As briefly noted at the end of Sec.~\ref{sec:role-multi-p}, this may require deriving an effective nonlocal Lagrangian $\mathcal{L}_{\rm eff}(\phi,\tilde{\phi})$ by introducing the composite field as an auxiliary field and integrating out the elementary field~$\chi(x)$. An effective-action approach to achieve this task will be pursued in~future~work.

\medskip

\noindent  \textbf{Ordinary resonances \textit{vs} ghost resonances.} To gain further insight into the complex conjugate pole structure of the dressed propagator, it is useful to make a detailed comparison between ordinary (unstable) resonances and ghost (anti-unstable) resonances. In particular, it would be interesting to formulate the QFT in a finite interval of time~\cite{Anselmi:2023phm,Anselmi:2023wjx} and compare the two systems for timescales much shorter than the inverse width, $t \ll 2/\Gamma$, to determine how the Breit-Wigner-like functions differ, and how these differences propagate into the opposite regime, $t \gtrsim 2/\Gamma$. This finite-time approach may also help strengthen our argument that the one-particle ghost state cannot be operationally isolated asymptotically, and will be studied~in~a~separate work.

\medskip

\noindent  \textbf{Complex poles and loop integrals.} An open question concerns loop integrals in modified perturbation theory, where the diagrammatic expansion is reorganized by replacing bare propagators with dressed ones and omitting additional self-energy insertions. Earlier work suggested that Lee-Wick-type contour deformations in loop integrals violate Lorentz invariance~\cite{Nakanishi:1971jj}, while more recent results have shown that Lorentz invariance can be preserved by suitably deforming also the spatial-momentum contour in the complex plane~\cite{Anselmi:2017yux,Anselmi:2025uzj}. It is therefore natural to ask whether similar complex deformations can be applied to loop integrals involving the dressed ghost propagator, consistently with both unitarity and Lorentz invariance. 

\medskip

\noindent  \textbf{Causality.} Different ghost quantizations can lead to different (micro)causality violations, for instance through an acausal classical limit~\cite{Anselmi:2018bra} or the presence of two microscopic arrows of causality~\cite{Donoghue:2019ecz,Aoki:2025uff}. While causality need not be a fundamental principle~\cite{Anselmi:2026uud}, our study so far finds no evidence of (micro)causality violation. First, our free propagators in perturbation theory are defined using the causal Feynman prescription: the usual convention is that positive (negative) energies propagate forward (backward) in time. Second, the dressed propagator also shows no acausal propagation once the absorptive contributions are properly identified~\cite{Buoninfante:2025klm}. Third, the quantum dynamics of the asymptotic ghost field discussed here appears causal: single arrow of causality and no faster-than-light signals. Nevertheless, further work is needed to determine whether first-sheet complex conjugate poles can induce  causality violations, for instance through loop effects in transition amplitudes~\cite{Grinstein:2008bg}. If so, it would then be interesting to understand whether the multi-particle masking mechanism can make such violations physically unobservable.

\medskip

\noindent  \textbf{Four-derivative theories.} Although our results are derived starting from the Lagrangian~\eqref{lagrangian}, we expect the analysis to be broadly general, provided the QFT framework and the locality assumption are maintained. In particular, a physically relevant application is to four-derivative field theories, where one propagator component is ordinary and massless, while the other is massive and ghost-like~\cite{Pais:1950za,Bender:2007wu,Salvio:2015gsi,Holdom:2023usn,Holdom:2024cfq,Lee:1969fy,Lee:1970iw}. In this case, the spectral representation of the four-derivative propagator above the multi-particle threshold and in the first Riemann sheet reads~\cite{Coleman:1969xz,Buoninfante:2025klm}
\begin{equation}
	\bar{G}_4(-p^2)=\frac{-i}{p^2-i\epsilon}+\frac{iZ}{p^2+M^2}+\frac{iZ^*}{p^2+M^{*2}}+\int_{M_{\rm th}^2}^\infty{\rm d}\sigma \frac{-i\rho(\sigma)}{p^2+\sigma-i\epsilon}\,,
	\label{spectral-four-deriv-examp}
\end{equation}
where $M_{\rm th}$ denotes the mass threshold for multi-particle production. Treating the complex conjugate poles and the associated asymptotic fields as done in this work, we can obtain a  local and Hermitian doubled-field description. The corresponding implications may be particularly relevant for quadratic gravity~\cite{Stelle:1976gc,Tomboulis:1980bs,Avramidi:1985ki,Salvio:2018crh,Anselmi:2018tmf,Donoghue:2021cza,Holdom:2021hlo,Buoninfante:2023ryt,Buoninfante:2025dgy,Kuntz:2024rzu,Oda:2025buc,Kumar:2026qnw} and hence quantum gravity~\cite{Buoninfante:2024yth,Basile:2024oms}, and will~be~studied~elsewhere.

\medskip

\noindent  \textbf{Analogy with (quasi-)dissipative systems.} It is possible to perform a change of basis from the operators $(\alpha_{\vec{p}},\beta_{\vec{p}})$ used in this work to new ones $(a_{\vec{p}},\tilde{a}_{\vec{p}})$, such that the real fields ($\eta$, $\tilde{\eta}$), and thus ($\phi^{\rm as},$ $\tilde{\phi}^{\rm as}$), admit mode decompositions in terms of the real frequency $\omega_{\vec{p}}=\sqrt{\vec{p}^2+m^2}$. This can be done by making the following Bogoliubov-like transformation~\cite{Nakanishi:1972wx}:
\begin{equation}
\left\lbrace\begin{array}{ll}	
	\alpha_{\vec{p}}\!\!\!&=\displaystyle  \frac{1}{(8\Omega_{\vec{p}}\,\omega_{\vec{p}})^{1/2}}\left[\big(\Omega_{\vec{p}}+\omega_{\vec{p}}\big)\big(a_{\vec{p}}+i\tilde{a}_{\vec{p}}\big)+\big(\Omega_{\vec{p}}-\omega_{\vec{p}}\big)\big(a_{-\vec{p}}^\dagger+i\tilde{a}^\dagger_{-\vec{p}}\big) \right]\\[4.5mm]
	\beta_{\vec{p}}\!\!\!&=\displaystyle  \frac{1}{(8\Omega^*_{\vec{p}}\,\omega_{\vec{p}})^{1/2}}\left[\big(\Omega^*_{\vec{p}}+\omega_{\vec{p}}\big)\big(a_{\vec{p}}-i\tilde{a}_{\vec{p}}\big)+\big(\Omega^*_{\vec{p}}-\omega_{\vec{p}}\big)\big(a_{-\vec{p}}^\dagger-i\tilde{a}^\dagger_{-\vec{p}}\big) \right]
	\end{array}
\right.\,,
\end{equation}
which preserves the canonical commutation relations and transforms the Hamiltonian~\eqref{hamilt-alphabeta} into
\begin{equation}
	H=\int {\rm d}^3p  \left[\omega_{\vec{p}}\left(\tilde{a}^{\dagger}_{\vec{p}}\,\tilde{a}_{\vec{p}}-a^\dagger_{\vec{p}}\,a_{\vec{p}}\right)+\frac{m\Gamma}{2\omega_{\vec{p}}}\left(\tilde{a}_{\vec{p}}\,a^\dagger_{\vec{p}}+\tilde{a}_{\vec{p}}^\dagger\, a_{\vec{p}}+\tilde{a}_{\vec{p}}\,a_{-\vec{p}}+\tilde{a}_{\vec{p}}^\dagger\, a_{-\vec{p}}^\dagger\right) \right]\,.
	\label{hamilt-aatilde}
\end{equation}
Without the last two terms and redefining $a_{\vec{p}}\rightarrow ia_{\vec{p}}^\dagger$, the expression~\eqref{hamilt-aatilde} reduces to the Hamiltonian of the field-theory generalization of a damped harmonic oscillator~\cite{Celeghini:1991yv}. This analogy is reinforced by the fact that a local Lagrangian description for a damped harmonic oscillator requires doubling the degrees of freedom (see~\cite{Celeghini:1991yv} and references therein). Moreover, the formalism used in~\cite{Celeghini:1991yv} is that of thermo field dynamics~\cite{Takahashi:1996zn,Umezawa:1993yq}, where thermal systems are described by doubling the field content: the untilded field represents the system and the tilded (thermal ghost) field the reservoir (or vice versa in our convention). In equilibrium the two sectors decouple, while out of equilibrium they are coupled, in analogy with~\eqref{hamilt-aatilde}. This raises the question of whether the interacting pair $(\phi^{\rm as}, \tilde{\phi}^{\rm as})$ can be interpreted as a quasi-dissipative system, i.e. whether the one-particle ghost state and the multi-particle component can be viewed as a system-reservoir pair (or vice versa).  Although the thermal ghost is typically quantized with negative energy and positive norm~\cite{Takahashi:1996zn,Umezawa:1993yq}, it is still interesting to investigate whether the above analogy can be~pushed~further. 
In fact, inspired by the effective doubling of degrees of freedom in thermo field dynamics, the doubling mechanism for the ghost field studied here -- according to which the quantum dynamics effectively increases by one the number of asymptotic fields -- can be named~\textit{dynamical~doubling}.

\medskip

Addressing the outlined open questions and applications may pave the way toward a deeper understanding of ghosts in (higher-derivative) QFT and shed new light on their true nature. All~these interesting aspects will be investigated in future works.


\subsection*{Acknowledgements}
I am grateful to John Donoghue, Bob Holdom, and Luca Smaldone for valuable feedback, and to Jisuke Kubo and Taichiro Kugo for helpful criticisms. 
I acknowledge financial support from the Xunta de Galicia (CIGUS Network), the EU through the Galicia Feder 2021-2027 Program, and the Grant CEX2023-001318-M funded by MICIU/AEI/10.13039/501100011033.



\bibliographystyle{utphys}
\bibliography{References}


\end{document}